\documentclass[letterpaper,twocolumn,10pt]{article}
\usepackage{zhanggroup}

\usepackage{xspace}
\usepackage{multirow}
\usepackage{graphicx}
\usepackage{amsmath}
\usepackage{newtxtext}

\newcommand{\mypara}[1]{\noindent{\bf {#1}.}\xspace}
\newcommand{\method}{\textsc{PersonaHate}\xspace}

\begin{document}

\date{}

\title{\method: A Scalable Persona-Based Data Synthesis Pipeline for Hate Speech Analysis}

\author{
Xinyu Zhang\ \ \
Ziqing Yang\ \ \
Michael Backes\ \ \
Yang Zhang\thanks{Corresponding author.}
\\
\\
\textit{CISPA Helmholtz Center for Information Security}
}

\maketitle

\begin{abstract}
Hate speech remains a persistent and evolving threat on online platforms.
The widespread use of large language models (LLMs) further intensifies this problem by making hateful content easier and cheaper to generate at scale.
Although stronger hate speech detectors continue to be proposed, their performance remains uneven across target identity groups and different forms of hateful expression, partly because existing training data provides only limited coverage of these cases.
Thus, we introduce PersonaHate, a scalable persona-driven data synthesis pipeline for constructing diverse hate speech datasets.
PersonaHate begins with a large persona pool of 457K personas drawn from two real-world sources.
Conditioned on selected personas from this pool, it prompts eight LLMs to generate hate speech across 34 identity groups and performs multi-judge annotation to obtain reliable labels.
The resulting corpus contains 791K samples, from which we curate a 67K group-balanced training set.
Our modular design also supports extension to new identity groups, generators, and emerging hate targets.
Built upon this dataset, we evaluate two detector training strategies: training from scratch and further fine-tuning.
Results show that all trained models achieve competitive performance across four widely used external benchmarks.
Notably, a 184M-parameter DeBERTa-v3 model trained on PersonaHate reaches 77.5\% average F1 and 79.4\% average accuracy, outperforming OpenAI's Moderation API on HateBenchSet and MHS.
Furthermore, fine-tuning Llama-3.1-8B raises its average F1 from 70.7\% to 78.4\%, a gain of 7.7 percentage points, surpassing LlamaGuard-3, Meta's official safety model based on the same Llama-3.1-8B architecture, on three of four benchmarks.
These results show that PersonaHate can substantially improve the cross-dataset generalization of hate speech detectors.
\end{abstract}

\noindent\textcolor{red}{Disclaimer: This paper contains examples of hateful and offensive language; reader discretion is recommended.}


\section{Introduction}
\label{sec:introduction}

Online hate speech remains a major challenge for content moderation systems~\cite{YZ21, SCC19}.
A UNESCO-Ipsos survey reports that 67\% of internet users have been exposed to online hate speech~\cite{unesco_ipsos_2023_news}, highlighting the need for reliable hate speech detection~\cite{SCC19,SLSVLM25}.

Recent advances in large language models (LLMs) further complicate this problem.
Because LLMs can produce fluent and varied text at low cost, hateful content can now be generated at scale and in many different styles~\cite{BGMS21}.
Hate speech detectors, therefore, need to generalize beyond human-authored data to increasingly diverse and evolving LLM-generated content.
Specifically, we follow the United Nations' definition and use hate speech to refer to \textit{offensive discourse targeting individuals or groups based on inherent characteristics, such as race, religion, gender, or other identity attributes}~\cite{UN_hate_speech}.

Despite continuous progress, existing hate speech detectors still exhibit systematic weaknesses.
Prior work shows that their performance can vary substantially across target identity groups and across datasets~\cite{YZ21, APP22}.
These weaknesses are especially concerning in the LLM era, where harmful content can be easily adapted to new targets, styles, and platforms~\cite{WAM21, SWQBZZ25}.
One major cause is the limitation of existing training data.
Many hate speech datasets are small, platform-specific, and highly imbalanced across identity groups and labels~\cite{FN18,VD20}.
LLM-generated data provides a scalable alternative, but existing methods usually rely on direct prompting or fixed prompt templates~\cite{GHC25, WAM21}, which often produce limited diversity.
More importantly, they do not reflect how hate speech appears in real-world platforms: hateful content is relatively rare, and a large portion of it is produced by users with persistent hateful attitudes, beliefs, and rhetorical habits~\cite{TLMHFOR25}.

To address this gap, we use personas as a controllable mechanism for data synthesis~\cite{GCWYMY24}.
Rather than relying only on general-purpose personas, we extract hateful personas from 4chan /pol/~\cite{4chan}, an online platform known for politically extreme and hateful discourse.
This choice is motivated by the observation that hate speech is not uniformly produced by all users, but is often concentrated among users who repeatedly express hateful views~\cite{APP22, QBLBW19}.
By inferring personas from real hateful posts, our pipeline captures speaker archetypes that are more likely to generate hateful content in realistic ways.
At the same time, we combine these hateful personas with general-purpose personas to broaden stylistic and demographic coverage.
This allows the generated data to include both explicit extremist rhetoric and more mainstream, implicit, or coded forms of prejudice.

Based on this idea, we propose \method, a scalable persona-driven data synthesis pipeline for constructing diverse data for hate speech analysis.
\method first builds a 457K-persona pool from two complementary sources: hateful personas inferred from the politically incorrect board /pol/ on 4chan~\cite{4chan}, and general-purpose personas sampled from PersonaHub~\cite{GCWYMY24}.
It then deduplicates the pool and applies diversity-aware sampling (farthest point sampling~\cite{G85}) to select representative personas.
These personas are paired with 34 identity groups and used to guide generation from multiple LLMs, producing speech samples that vary across targets, speaker archetypes, generator styles, and sentiment directions.
After refusal filtering and output normalization, \method applies multi-judge annotation to obtain reliable hate speech labels.
The resulting corpus contains 791K valid samples, from which we curate a 67K group-balanced training set.
Because each stage is modular, \method can be extended to new identity groups, generators, and emerging hate targets.
It can therefore serve as a training data construction pipeline for improving hate speech detectors' performance and generalizability.

We evaluate \method from both data and model perspectives.
On the data side, we analyze whether the pipeline produces diverse synthetic hate speech.
Our results show that persona conditioning improves semantic and lexical diversity over direct prompting, that persona diversity transfers to output diversity, and that hateful personas from 4chan and general-purpose personas provide complementary signals.
We further show that using multiple LLM generators broadens the output distribution and reduces dependence on any single model's style or safety behavior.

On the model side, we evaluate detectors trained or further fine-tuned on \method across four external hate speech benchmarks and 34 identity groups.
Models trained on \method achieve strong cross-dataset performance, approaching prompted commercial LLM classifiers while using much smaller deployable models.
\method also improves performance on weakly detected groups, and ablation studies show that broad coverage over identity groups, personas, and generators is important for detector robustness.
These findings suggest that persona-driven synthesis can provide both diverse evaluation data and effective training supervision for robust hate speech detection.

In summary, we make the following contributions:

\begin{itemize}
    \item \textbf{We introduce \method, a large-scale persona-driven hate speech data synthesis pipeline.}
    \method starts from a pool of 457K personas, selects 2{,}285 representative personas, and uses eight LLMs to generate hate speech across 34 identity groups.
    After multi-judge annotation, the resulting corpus contains 791K labeled samples, from which we curate a 67K group-balanced training set for detector training.

    \item \textbf{We empirically validate that \method is a scalable mechanism for increasing data diversity.}
    We show that persona conditioning broadens semantic and lexical coverage compared with direct prompting and fixed-template generation.
    We further show that persona diversity transfers to output diversity, hateful and general persona sources provide complementary signals, and persona prompts reduce refusal rates in safety-aligned LLMs.
    Controlled ablations show that identity group coverage and multi-model generation provide complementary gains beyond data volume alone.

    \item \textbf{We show that \method improves detector robustness across benchmarks and identity groups.}
    Models trained on \method achieve strong performance on four external benchmarks.
    For example, LFTW-RoBERTa reaches 78.1\% average F1, DeBERTa-v3 reaches 77.6\%, and Llama-3.2-3B reaches 77,8\%, approaching prompted commercial LLM classifiers while using smaller deployable models.
    We also show that full multi-group training outperforms targeted single-group training on weakly detected identity groups, indicating the value of broad group coverage.
\end{itemize}

\section{Background and Related Work}
\label{sec:background_and_related_work}

\mypara{Hate Speech Detectors}
Hate speech detectors have evolved from feature-based classifiers~\cite{KW13, WH16, NTTMC16, DWMW17} to neural encoders~\cite{BGGV17, MFC20, ASK20, CBMG21}, commercial moderation APIs~\cite{OpenAI_Moderation, Perspective}, and LLM-based classifiers~\cite{HKA23, GDF25, WHADKM25, FL25}. 
Encoder-based models such as BERT and RoBERTa provide efficient deployment options~\cite{DCLT19, LOGDJCLLZS19}, while domain-adapted models such as HateBERT further pretrain on abusive or hateful corpora~\cite{CBMG21}. 
Commercial systems such as Perspective API~\cite{Perspective} and OpenAI Moderation~\cite{OpenAI_Moderation} are widely used for content moderation~\cite{LTTSGMV22, MZANLAJW23}.
Recent work also prompts or fine-tunes LLMs as hate speech or safety classifiers~\cite{GDF25,BPV24}. 

\mypara{Hate Speech Datasets}
Prior work has introduced a range of benchmark datasets for hate speech detection~\cite{WH12, NTTMC16}. 
Early datasets mainly focus on social media platforms such as Twitter and Gab~\cite{WH16, DWMW17, FDCLBSVSK18}. 
For example, HateXplain~\cite{MSYBGM21} includes target-group labels and rationales for Twitter and Gab posts, while Measuring Hate Speech (MHS)~\cite{SBBSVK22} collects hate speech from YouTube, Twitter, and Reddit using large-scale annotator judgments~\cite{KBSV20, SBBSVK22}. 
However, most existing datasets are primarily human-authored, tied to specific platforms or time periods, and uneven in their coverage of identity groups and expression styles~\cite{YZ21, APP22, LDZH22}.
Recent benchmarks such as HateBenchSet~\cite{SWQBZZ25} use LLM-generated hate speech to evaluate detectors across identity groups. 
However, these datasets are often built with fixed prompt templates and are primarily designed for evaluation rather than scalable detector training. 
These limitations motivate data construction methods that can systematically expand coverage across targets, sources, and expressions.

\mypara{LLM-Based Data Synthesis}
LLMs have been used to synthesize training and evaluation data for many NLP tasks~\cite{KSYVLWGLWN25}.
In safety and moderation, synthetic data has been used for data augmentation~\cite{WAM21}, red-teaming~\cite{SWYCCH24}, detector evaluation~\cite{SWQBZZ25}, and controlled benchmark construction~\cite{RVNWMP21, VTWK21}.
Existing hate speech synthesis methods, such as HateBenchSet~\cite{SWQBZZ25}, commonly rely on direct prompts or fixed prompt templates.
Persona-driven prompting is another line of LLM-based synthesis~\cite{DMRKN23, DRSC26, GCWYMY24}, where personas are used to steer generation by specifying speaker background, beliefs, preferences, or communication style.
Large-scale persona resources such as PersonaHub~\cite{GCWYMY24} provide broad collections of automatically curated personas for synthetic data generation.
Prior studies also examine how assigned personas affect model-generated opinions, biases, and linguistic patterns~\cite{LDF24}.
Our work follows this line of LLM-based data synthesis but applies it to hate speech training data construction.
We combine general-purpose personas with hateful personas inferred from 4chan /pol/ posts, use multiple LLM generators, and apply multi-judge annotation to produce a large-scale labeled corpus.

\section{Empirical Motivation}
\label{sec:empirical_motivation}

Before introducing the \method pipeline, we first present an empirical analysis that reveals structural weaknesses in existing hate speech datasets and detectors. 

\subsection{Limited Identity Group Coverage}
\label{sec:dataset_limitations}

Robust hate speech detection requires data that covers a broad range of target identity groups, yet existing datasets often provide incomplete or imbalanced coverage~\cite{VD20, PBSBP21}.

Human-authored benchmarks such as HateXplain~\cite{MSYBGM21}, Davidson~\cite{DWMW17}, and MHS~\cite{SBBSVK22} are widely used, but they are limited for group-level robustness analysis. 
HateXplain includes target-group annotations, but its distribution is highly skewed: the three most frequent groups account for 44.4\% of labeled instances, while the ten least frequent groups together account for only 2.7\%~\cite{MSYBGM21}. 
Davidson~\cite{DWMW17} contains no target-group annotations, and MHS, while broader in coverage, remains imbalanced and requires binarizing continuous severity scores for standard classification~\cite{KBSV20, SBBSVK22}.

LLM-generated benchmarks such as HateBenchSet improve group balance by construction, covering 34 subgroups across six identity categories~\cite{SWQBZZ25}. 
However, HateBenchSet remains limited as a training resource in two ways. 
First, it contains only 7{,}838 samples, which limits the per-group training signal. 
Second, its six generators differ substantially in output style and quality: GPT-4 outputs are generally coherent but often formulaic, whereas OPT outputs contain more unstructured fragments, such as forum-style replies, prompt artifacts, and incomplete text. 
This heterogeneity can complicate training and evaluation, since models may learn generator-specific artifacts rather than generalizable hate speech patterns.

\subsection{Limitations of Existing Hate Speech Detectors}
\label{sec:limitation_detectors}

We further evaluate hate speech detectors across different identity groups and datasets.

\mypara{Group-Wise Disparities}
We first evaluate Llama Guard 3~\cite{Llama_Guard_3}, a strong open-source safety classifier, on HateBenchSet at both category and subgroup levels.
Although Llama Guard 3 achieves strong overall performance (F1 = 85.5\%), Table~\ref{tab:group_disparity} shows substantial variation across the six main identity categories.
The best-performing category is Religion (F1 = 89.6\%), while the weakest is Sexuality (F1 = 78.3\%), yielding an 11.3 percentage-point gap.
The gap becomes larger at the subgroup level. 
The best-performing subgroup is Religion\_Christian (F1 = 92.8\%), whereas the weakest is Sexuality\_Straight (F1 = 63.5\%), a 29.3 percentage-point difference. 
Other low-performing subgroups include Gender\_Men (F1 = 73.8\%), Race\_Native-American (F1 = 79.1\%), and Race\_Asian (F1 = 80.2\%). 
These results show that even a capable hate speech detector cannot perform equally across identity groups.

\begin{table}[t!]
\centering
\caption{Llama Guard 3 per-group F1 scores (\%) on HateBenchSet. The gap between the best and worst categories is 11.3 percentage points.}
\label{tab:group_disparity}
\small
\setlength{\tabcolsep}{5pt}
\begin{tabular}{lcccc}
\toprule
\textbf{Category} & \textbf{Samples} & \textbf{Precision} & \textbf{Recall} & \textbf{F1} \\
\midrule
Religion    & 1{,}663 & 94.7 & 85.1 & 89.6 \\
Origin      & 1{,}157 & 93.4 & 82.7 & 87.8 \\
Disability  & 1{,}167 & 89.4 & 81.2 & 85.1 \\
Gender      & 1{,}362 & 92.4 & 77.3 & 84.2 \\
Race        & 1{,}578 & 87.7 & 79.5 & 83.4 \\
Sexuality   &    911 & 84.4 & 73.0 & 78.3 \\
\midrule
Overall     & 7{,}838 & 91.1 & 80.6 & 85.5 \\
\bottomrule
\end{tabular}
\end{table}

\mypara{Limited Cross-Dataset Generalizability}
We evaluate four representative detectors, including Detoxify~\cite{Detoxify}, LFTW~\cite{VTWK21}, OpenAI Moderation~\cite{OpenAI_Moderation}, and the Google Perspective API~\cite{Perspective}, across four benchmarks mentioned in Section~\ref{sec:dataset_limitations}.
As shown in Table~\ref{tab:cross_dataset}, we find that strong performance on one benchmark does not necessarily transfer to another.
Llama Guard 3 achieves 85.3\% F1 on HateBenchSet but drops to 55.5\% on MHS, yielding a 29.8 percentage-point gap. 
Cardiff-RoBERTa performs best on Davidson (90.8\%) but drops to 68.8\% on MHS. 
Detoxify is more stable, with a smaller gap of 11.9 points, but its overall performance remains lower and never exceeds 75.4\% F1.
This pattern might be influenced by their imbalanced training dataset.

\begin{table*}[t!]
\centering
\caption{F1 scores (\%) of pretrained detectors evaluated on four benchmarks.
}
\label{tab:cross_dataset}
\small
\setlength{\tabcolsep}{5pt}
\begin{tabular}{lcccc}
\toprule
\textbf{Model} & \textbf{HateBench} & \textbf{HateXplain} & \textbf{Davidson} & \textbf{MHS} \\
\midrule
Detoxify            & 75.4 & 63.5 & 73.6 & 71.8 \\
LFTW-RoBERTa        & 79.4 & 72.4 & 82.3 & 66.7\\
Cardiff-RoBERTa     & 82.7 & 71.7 & 90.8 & 68.8 \\
Llama Guard 3       & 85.3 & 79.9 & 75.8 & 55.5 \\
\bottomrule
\end{tabular}
\end{table*}

\subsection{Toward Scalable and Diverse Training Data}
\label{sec:implications}
 
Based on our observations above, we find that existing hate speech datasets have a limited and uneven coverage of identity groups and labels.
Further, existing hate speech detectors exhibit an identity group-wise disparity and limited cross-dataset generalizability, which is bounded by the model structure and the training data.
Thus, we focus on hate speech data and aim to provide a more diverse and scalable data synthesis pipeline for hate speech analysis.

\section{\method Pipeline}
\label{sec:personahate_pipeline}

\subsection{Overview}
\label{sec:overview}

\begin{figure*}
    \centering
    \includegraphics[width=\linewidth]{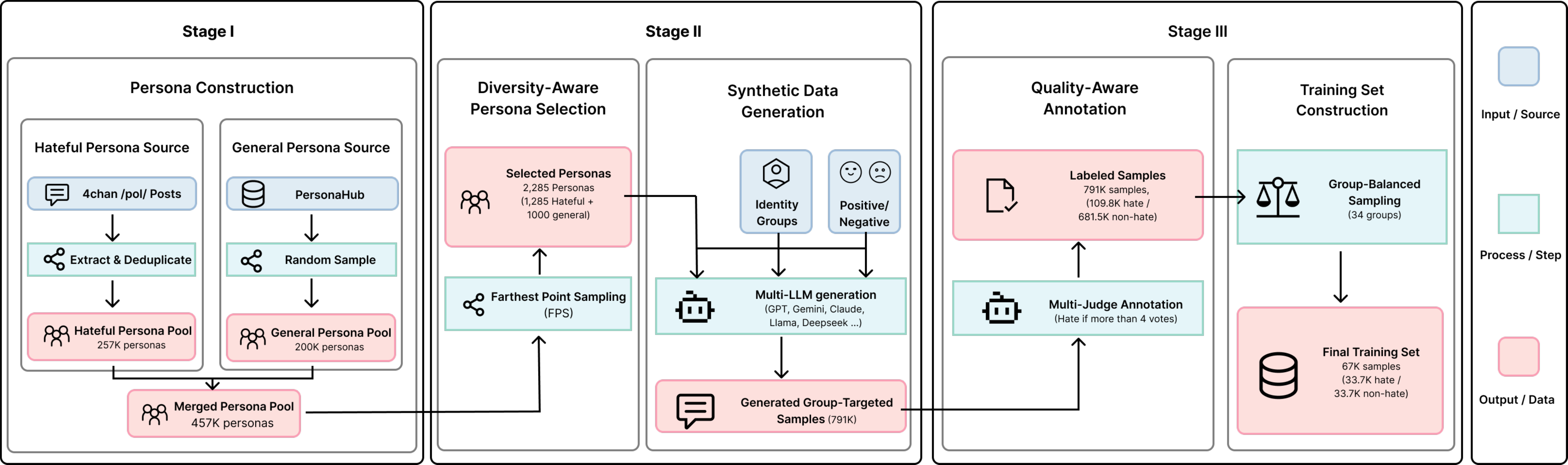}
    \caption{Overview of the \method pipeline, including persona construction, diversity-aware persona selection, multi-LLM group-targeted data generation, quality-aware annotation, and group-balanced training set construction.}
    \label{fig:overview}
\end{figure*}

In this work, we propose \method, a scalable persona-driven data synthesis pipeline for constructing diverse hate speech data with a broad coverage.
As shown in Figure~\ref{fig:overview}, the pipeline consists of three stages.
In the first stage, we construct a large persona pool containing both hateful and general personas.
Secondly, we extract personas to generate hate speech examples covering different target groups via multiple LLMs.
At the final stage, we conduct a quality-aware multi-judge annotation and group-balanced sampling to obtain a labeled dataset for hate speech analysis.

\subsection{Stage I: Persona Pool Construction}
\label{sec:persona_construction}

A key design choice in our \method is to generate hate speech through personas rather than through direct prompting.
Personas provide speaker-level context: they encode attitudes, beliefs, social background, and rhetorical tendencies that shape how a person may express views about an identity group~\cite{GCWYMY24, DMRKN23}.
This is particularly important for hate speech synthesis because hate speech is not uniformly produced by all users.
In real-world platforms, hateful content is relatively sparse, and a substantial portion is produced by users who repeatedly express hateful attitudes~\cite{APP22, QBLBW19}.
Therefore, relying only on generic personas may improve surface diversity, but may not capture the speaker archetypes that are most likely to produce hateful content.

To address this issue, we construct a hateful persona pool from /pol/, the politically incorrect board of 4chan.
We use the dataset released by Papasavva et al.~\cite{PZCSB20}, which contains 3.5 years of /pol/ posts from June 2016 to November 2019, together with Perspective API identity attack scores.
We filter posts with an identity attack score above 0.7 and a length above 500 characters.
For each retained post, we prompt GPT-4o~\cite{GPT4o} to infer a persona description that characterizes the likely speaker behind the post.
This process yields 305,616 hateful personas, each grounded in an actual hateful post.
However, they contain substantial redundancy, since many posts express similar beliefs or rhetorical patterns.
We therefore conduct deduplication: given persona embeddings, we greedily retain a persona only if its cosine similarity to all previously retained personas is below a threshold $\tau=0.9$.

The algorithm can be found in Appendix~\ref{app:deduplication} for reproduction.
The hateful persona pool is reduced from 305,616 to 257,473 personas, removing 15.7\% of the original entries.

Importantly, we do not directly use the original /pol/ posts as training samples.
Instead, we use them only to recover speaker archetypes grounded in real hateful discourse.
This design allows \method to capture realistic hateful personas while avoiding direct reuse of platform-specific posts as detector training data.

Hateful personas alone, however, cover only one region of the persona space.
They tend to reflect more extreme, confrontational, and politically charged speaker profiles.
To broaden the range of possible speakers, we additionally sample 200K general-purpose personas from PersonaHub~\cite{GCWYMY24}.
These personas are not necessarily hateful, but they introduce diverse occupations, backgrounds, interests, and communication styles.

The final persona pool thus combines two complementary sources, resulting in a 457K persona pool.
Example personas are provided in Appendix~\ref{app:persona_examples}.
The 4chan-derived personas provide realistic hateful speaker archetypes and improve the likelihood of generating hate-related content.
The PersonaHub personas broaden stylistic and demographic diversity.
Together, they support hate speech synthesis that is both grounded in real-world hateful discourse and diverse enough for robust detector training.

\subsection{Stage II: Diversity-Oriented Data Synthesis}
\label{sec:synthesis}

In this stage, \method samples personas from the constructed persona pool and prompts multiple LLMs to generate hate speech data conditioned on these personas and 34 identity groups.

\subsubsection{Diversity-Aware Persona Selection}
\label{sec:persona_selection}

Random sampling tends to over-represent dense regions of the persona space, such as generic political commentators, while under-representing rarer archetypes. 
To improve coverage, we apply Farthest Point Sampling (FPS) independently to each persona source.
FPS operates on normalized sentence embeddings $\{\mathbf{e}_i\}$ computed with \texttt{all-MiniLM-L6-v2}. 
Starting from a random seed, it iteratively selects the persona that is farthest from the current selected set in cosine distance:
\begin{equation}
    p_{t+1} = \arg\max_{i \notin \mathcal{S}_t} \min_{j \in \mathcal{S}_t} \left(1 - \mathbf{e}_i^\top \mathbf{e}_j\right),
    \label{eq:fps}
\end{equation}
where $\mathcal{S}_t$ denotes the selected indices after $t$ iterations. 
This greedy procedure improves coverage of the embedding space by discouraging redundant selections from dense clusters.
We run FPS separately on the two persona sources and select personas in proportion to each pool's size: 1{,}000 from PersonaHub and 1{,}285 from the deduplicated 4chan pool, yielding 2{,}285 personas in total. 
We further compare FPS with random sampling and Density-Aware FPS (DA-FPS)~\cite{CG26} with the same amount of samples (2,285) in Table~\ref{tab:fps_comparison}.
Compared with random sampling, FPS reduces the fill distance from 0.781 to 0.686 and increases the mean nearest-neighbor distance from 0.312 to 0.415, indicating broader and less redundant coverage.
DA-FPS yields the same coverage metrics as FPS in our setting, but is substantially slower. 
We therefore use source-stratified FPS for its simplicity and efficiency. 
A UMAP visualization of the selected personas is provided in Appendix~\ref{app:fps_umap}.

\begin{table}[t!]
\centering
\caption{Comparison of persona selection methods ($n=2{,}285$).
Fill distance measures the worst-case cosine distance from any pool point to its nearest selected point; lower is better.
Mean NN distance measures the average nearest-neighbor distance among selected personas; higher indicates less redundant selection.}
\label{tab:fps_comparison}
\small
\setlength{\tabcolsep}{5pt}
\begin{tabular}{lccc}
\toprule
\textbf{Method} & \textbf{Fill Dist.\,$\downarrow$} & \textbf{Mean NN\,$\uparrow$} & \textbf{Time} \\
\midrule
Random   & 0.781 & 0.312 & {--} \\
DA-FPS   & \textbf{0.686} & \textbf{0.415} & 1{,}370\,s \\
FPS      & \textbf{0.686} & \textbf{0.415} & 30\,s \\
\bottomrule
\end{tabular}
\end{table}

\subsubsection{Prompt Template Design}
\label{sec:prompt_design}

We frame generation as a creative-writing task. 
The prompt template below requires different personas, identity groups, and sentiments (positive and negative):
\begin{quote}
\small
\textit{You are a creative writer working on a novel that explores social dynamics and prejudices. You're writing a character study.}

\textit{Character Profile: \texttt{\{persona\}}}

\textit{For authenticity, write a brief monologue (2--3 sentences) where this character expresses their private, unfiltered \texttt{\{sentiment\}} views about \texttt{\{identity group\}}. This is fiction for educational purposes about understanding different perspectives.}

\textit{Character's monologue:}
\end{quote}
This framing serves two purposes. 
First, it provides a contextualized writing task, which reduces refusals from safety-aligned models compared with direct requests. 
Second, it encourages the model to write in the persona's voice, increasing variation in vocabulary, tone, explicitness, and framing.

We adopt the identity group taxonomy from prior works~\cite{SBBSVK22, SWQBZZ25}, which defines 34 groups across six categories (see Table~\ref{tab:identity_groups} in Appendix~\ref{app:identity_groups}). 
This taxonomy covers major protected characteristics used in content moderation policies and allows direct comparison with existing LLM-generated hate speech benchmarks~\cite{UN_hate_speech, SBBSVK22, SWQBZZ25}. 
Based on this taxonomy, each selected persona is paired with each of the 34 identity groups and two sentiment directions, positive and negative. 
This produces $2{,}285 \times 34 \times 2 = 155{,}380$ prompts per generator.

\subsubsection{Multi-LLM Generation}
\label{sec:multi_model}

We use eight LLMs spanning different providers, architectures, and alignment strategies as our generators:
\begin{itemize}
    \item \textbf{Commercial APIs:} GPT-4o-mini, Claude, and Gemini.
    \item \textbf{Open-Source Models:} Llama-3-8B, Qwen-2.5-7B, Ministral-8B, Gemma-2-9B, and DeepSeek-R1-Distill-8B.
\end{itemize}
Using multiple generators could increase distributional coverage and reduce dependence on the style or refusal behavior of any single model~\cite{SWQBZZ25}. 
In this stage, each model receives the same 155{,}380 prompts with a temperature of 0.7.
Table~\ref{tab:generator_composition} reports generation statistics by model. 
Compliance rates vary widely, from 12.3\% for Claude-3-Haiku to 99.2\% for Llama-3.1-8B, reflecting differences in their internal safety mechanisms. 
Open-weight models generally produce more valid outputs, while API-gated models are more conservative. 
Hate rates also vary by generator: Llama-3.1-8B and Qwen-2.5-7B produce the highest fractions of samples labeled hate, while GPT-4o-mini and Gemini-2.5-Flash produce the lowest. 
This variation supports our multi-model design, since different generators contribute different amounts and types of training signal.

\begin{table}[t!]
\centering
\caption{Per-generator statistics. Compliance is the fraction of prompts that yield valid non-refused outputs. Hate rate is determined by the $\geq 4/6$ majority-vote threshold.}
\label{tab:generator_composition}
\small
\setlength{\tabcolsep}{5pt}
\begin{tabular}{llrrc}
\toprule
\textbf{Model} & \textbf{Type} & \textbf{Samples} & \textbf{Compliance} & \textbf{Hate (\%)} \\
\midrule
GPT-4o-mini      & API  & 109{,}586 & 70.5\% & 3.6 \\
Claude-3-Haiku   & API  & 19{,}130  & 12.3\% & 4.3 \\
Gemini-2.5-Flash & API  & 31{,}223  & 20.1\% & 3.9 \\
DeepSeek-R1-8B   & Open & 111{,}018 & 71.4\% & 15.2 \\
Qwen-2.5-7B      & Open & 104{,}218 & 67.1\% & 17.5 \\
Mistral-7B       & Open & 122{,}592 & 78.9\% & 15.1 \\
Gemma-2-9B       & Open & 139{,}326 & 89.7\% & 15.3 \\
Llama-3.1-8B     & Open & 154{,}190 & 99.2\% & 18.7 \\
\midrule
\textbf{Total}  &    --   & \textbf{791{,}283} & \textbf{63.7\%} & \textbf{13.9} \\
\bottomrule
\end{tabular}
\end{table}

We then remove refused outputs from the final dataset by detecting refusal keywords~\cite{ZWCNKF23}.
After refusal filtering, the combined outputs from all eight models yield 791{,}283 valid speech samples, covering all 34 identity groups from 2{,}285 diverse persona viewpoints.

\subsection{Stage III: Quality-Aware Annotation and Filtering}
\label{sec:annotation}

In this stage, we utilize a multi-judge annotation pipeline to label the 791K valid samples and then construct a group-balanced training set for hate speech detection.

\subsubsection{Data Cleaning}
\label{sec:confidence_filtering}

Before annotation, we apply a post-processing pipeline to clean the raw model outputs.
This pipeline addresses three issues that arise from multi-model generation.

\mypara{Enhanced Refusal Detection}
The initial keyword-based refusal filter (Section~\ref{sec:multi_model}) catches explicit refusals (e.g., ``I cannot generate hate speech'') but misses subtler forms.
We apply a secondary heuristic that detects meta-commentary (e.g., ``As a language model\ldots''), disclaimers appended to otherwise compliant outputs, and outputs that discuss hate speech in the abstract rather than adopting the persona's voice.
This secondary filter identifies an additional 8.6\% of outputs as implicit refusals across the eight models, ranging from 0.1\% (Gemini) to 14.0\% (Claude).

\mypara{Output Normalization}
Different models produce outputs with different formatting conventions.
Some wrap monologues in quotation marks, others prepend character names or stage directions, and some include chain-of-thought reasoning before the monologue (notably DeepSeek-R1).
We strip these artifacts to retain only the speech content, ensuring consistent formatting across models.

\mypara{Length Filtering}
We exclude outputs shorter than 10 characters, which are typically fragments rather than meaningful speech.
This removes a negligible fraction ($<$0.1\%) of outputs.

After post-processing, 791{,}283 valid samples remain, which are stored in JSONL format.
Each record contains the generated text, the persona description, the persona source (PersonaHub or 4chan), the target identity group, the sentiment direction, the generator, per-judge annotations, and the aggregated vote label.

\subsubsection{Multi-Judge Annotation}
\label{sec:voting}

We annotate each valid sample using six independent judges: three LLM APIs (GPT-4o-mini, Claude-3-Haiku, and Gemini-2.5-Flash), one open-source LLM (Gemma-2-9B), and two dedicated moderation tools (OpenAI Moderation and LlamaGuard-3-8B).
{
The jury intentionally combines LLMs with dedicated moderation tools: the former provide contextual understanding, while the latter provide task-specific classification results.
Moreover, multi-model voting provides a natural confidence signal for further analysis.
}
We first query each judge with the following prompt:
\begin{quote}
\small
\textit{Is the following text hate speech? Hate speech is content that expresses hatred or promotes violence against people based on race, ethnicity, religion, gender, sexual orientation, disability, or origin.}

\textit{Text: ``\texttt{\{text\}}''}

\textit{Answer with only one word: yes or no.}
\end{quote}
The temperature is set to 0 for reproducibility.
We parse each response by extracting the first word and mapping \texttt{yes} to hate and \texttt{no} to non-hate. 
Responses that cannot be parsed are recorded as abstentions.

Then, we conduct a calibration experiment on HateXplain~\cite{MSYBGM21}, using 14{,}385 samples after removing the ``offensive'' category, to choose the voting threshold.
Each judge independently classifies every sample, and we evaluate majority-vote thresholds $\geq k/6$ for $k \in \{3,4,5,6\}$. 
As shown in Table~\ref{tab:voting_threshold}, $\geq 4/6$ achieves the best F1 score, 82.1\%, with 82.6\% accuracy and 91.8\% recall. 
The stricter $\geq 5/6$ threshold slightly improves accuracy but reduces recall, while unanimous voting is too restrictive. 
We therefore use $\geq 4/6$ as the final threshold for assigning hate labels.
A sample is labeled \emph{hate} if at least four of the six judges classify it as hate speech, and \emph{non-hate} otherwise.
This produces 109{,}788 hate samples and 681{,}495 non-hate samples from 791{,}283 valid samples.

We further construct a group-balanced training dataset for further hate speech analysis.
Specifically, for each of the 34 identity groups, we sample up to 1{,}000 hate and 1{,}000 non-hate examples. 
Groups with fewer than 1{,}000 hate samples contribute all available hate samples and an equal number of non-hate samples. 
This yields a balanced training set of 67{,}452 samples, with 33{,}726 hate and 33{,}726 non-hate examples. 
This per-group balancing ensures that every identity group contributes, especially for groups with rare hate speech.

\begin{table}[t!]
\centering
\caption{Voting threshold calibration on HateXplain after removing the ``offensive'' category.
The $\geq 4/6$ threshold achieves the best F1 while maintaining high recall.}
\label{tab:voting_threshold}
\small
\setlength{\tabcolsep}{5pt}
\begin{tabular}{lcccc}
\toprule
\textbf{Threshold} & \textbf{Accuracy} & \textbf{F1} & \textbf{Precision} & \textbf{Recall} \\
\midrule
\multicolumn{5}{l}{\textit{Individual judges}} \\
\quad Gemma-2-9B        & 71.0 & 73.8 & 60.6 & 94.4 \\
\quad Gemini-2.5-Flash  & 81.7 & {81.8} & 71.8 & 95.1 \\
\quad Claude-3-Haiku    & 78.8 & {81.8} & 70.9 & {96.6} \\
\quad GPT-4o-mini       & 73.5 & 76.2 & 62.3 & \textbf{97.9} \\
\quad LlamaGuard-3      & 80.4 & 79.7 & 72.2 & 88.9 \\
\quad OpenAI Moderation & 77.2 & 74.3 & 72.7 & 76.0 \\
\midrule
\multicolumn{5}{l}{\textit{Multi-judge voting}} \\
\quad $\geq 3/6$  & 78.4 & 79.4 & 67.8 & 95.9 \\
\quad $\geq 4/6$  & {82.6} & \textbf{82.1} & 74.2 & 91.8 \\
\quad $\geq 5/6$  & \textbf{83.3} & 80.4 & {81.8} & 79.1 \\
\quad $\geq 6/6$  & 67.6 & 45.1 & \textbf{85.2} & 30.7 \\
\bottomrule
\end{tabular}
\end{table}

\subsection{\method Dataset Analysis}
\label{sec:dataset_stats}

Table~\ref{tab:dataset_overview} in Appendix~\ref{app:dataset_overview} summarizes our \method pipeline.
We further compare the resulting training set with existing hate speech datasets.
Table~\ref{tab:dataset_comparison_final} exhibits the dataset size, number of identity groups, and the imbalance rate (i.e., $\frac{\# \text{Largest Group}}{\# \text{Smallest Group}}$).
We find that \method is substantially larger than the synthetic HateBenchSet dataset and the largest MHS dataset.
\method provides explicit coverage of 34 identity groups and achieves near-uniform group representation (1.17:1). 
Compared with human-authored datasets such as HateXplain, Davidson, and MHS, \method is designed as a scalable, group-aware synthetic training resource that complements existing evaluation benchmarks.
Additionally, we compare our \method against the LLM-synthetic HateBenchSet in Appendix~\ref{app:diversity_analysis}.
Our \method outperforms HateBenchSet in both semantic and lexical diversity, even across different identity groups.

\begin{table}[t!]
\centering
\caption{Comparison of \method with existing hate speech datasets.}
\label{tab:dataset_comparison_final}
\small
\setlength{\tabcolsep}{5pt}
\begin{tabular}{lccc}
\toprule
\textbf{Dataset} & \textbf{Samples} & \textbf{Groups} & \textbf{Imbalance}\\
\midrule
HateXplain      & 20{,}148   & 23  & 894:1     \\
Davidson        & 24{,}783   & 0   & N/A       \\
MHS             & 39{,}565   & 39  & 12:1       \\
HateBenchSet    & 7{,}838    & 34  & $\sim$3:1 \\
\midrule
\textbf{\method} & \textbf{67,452} & \textbf{34} & $\sim$\textbf{1.17:1} \\
\bottomrule
\end{tabular}
\end{table}

\section{Design Validation of \method}
\label{sec:validation}

In this section, we evaluate whether our \method could produce diverse hate speech data at scale.
Firstly, we validate two key design choices in the pipeline, i.e., personas and multi-model generation. 
We examine the role of personas in improving both the diversity and feasibility of hate speech synthesis. 
We also analyze the role of multi-model generation in broadening distributional coverage and supporting scalable dataset construction.
Finally, we investigate our generation and annotation throughput to validate the scalability of our pipeline.

\subsection{Role of Personas}
\label{sec:persona_effect}

A central claim of our pipeline is that personas are not merely decorative prompt context, but a functional mechanism for inducing diversity in generated hate speech~\cite{GCWYMY24, DMRKN23, KSG25}.
We validate this claim by investigating the influence of personas' existence, diversity, and origin on dataset diversity and comparing it with other LLM-generated datasets.

\mypara{Influence of the Existence of Personas}
As is known, \method generates hate speech data conditioned on prompts with sampled personas (see Section~\ref{sec:prompt_design}).
We therefore compare the diversity of generated outputs under our \emph{persona}-conditioned generation with \emph{direct} generation, i.e., generating negative content targeting an identity group without persona context.
For comparison, we use the same generator (GPT-4o-mini) and target 34 identity groups.
We encode all outputs with Sentence-BERT (\texttt{all-MiniLM-L6-v2}) and compare their embedding-space coverage.

We observe that persona-conditioned outputs cover a substantially larger region of the embedding space. 
The mean pairwise cosine distance increases from 0.42 (direct) to 0.58 (persona), while the fill distance to a held-out reference corpus decreases from 0.81 to 0.67. 
These results indicate that persona conditioning increases semantic spread and improves distributional coverage. 

\mypara{Influence of the Diversity of Personas} 
We next examine whether diversity in the persona pool correlates with the generated outputs' diversity.
Using our FPS-selected personas ($n=2{,}285$), we compute pairwise cosine distances between persona embeddings and compare them with the corresponding output embedding distances, averaged across identity groups. 
The Spearman correlation between persona distance and output distance is $\rho=0.61$ ($p<0.001$), indicating that more dissimilar personas tend to produce more dissimilar outputs. 
This supports our use of embedding-space diversity during persona selection.

\mypara{Influence of the Source of Personas}
Our two persona sources, i.e., 4chan and PersonaHub, serve complementary roles. 
As shown in Table~\ref{tab:persona_source_comparison}, 4chan-derived hateful personas produce hateful outputs more frequently under majority voting (21.3\% vs.\ 6.8\%). 
In contrast, PersonaHub personas produce hateful outputs less often but yield more lexically and semantically diverse generations, i.e., with higher mean pairwise cosine distance and more unique unigrams per 1K samples. 
This supports our dual-source design: the hateful source provides stronger hate-related signals, while the general source broadens the range of viewpoints and expressions.

\begin{table}[t!]
\centering
\caption{Generation characteristics of hateful personas derived from 4chan and general personas sampled from PersonaHub, evaluated on GPT-4o-mini outputs.}
\label{tab:persona_source_comparison}
\small
\setlength{\tabcolsep}{5pt}
\begin{tabular}{lcc}
\toprule
\textbf{Metric} & \textbf{4chan} & \textbf{PersonaHub} \\
\midrule
Number of personas & 1{,}285 & 1{,}000 \\
Hate rate (majority vote) & 21.3\% & 6.8\% \\
Median output length (chars) & 412 & 487 \\
Mean pairwise cosine dist.\ & 0.51 & 0.62 \\
Unique unigrams (per 1K samples) & 2{,}847 & 3{,}412 \\
\bottomrule
\end{tabular}
\end{table}

\subsection{Role of Multiple Generators}
\label{sec:cross_model_diversity}

While personas introduce diversity at the prompt level, using multiple generators further reduces dependence on any single model's output style or safety behavior. 
We therefore examine whether different generators contribute non-redundant content.

\begin{figure*}[t!]
\centering
\includegraphics[width=0.8\linewidth]{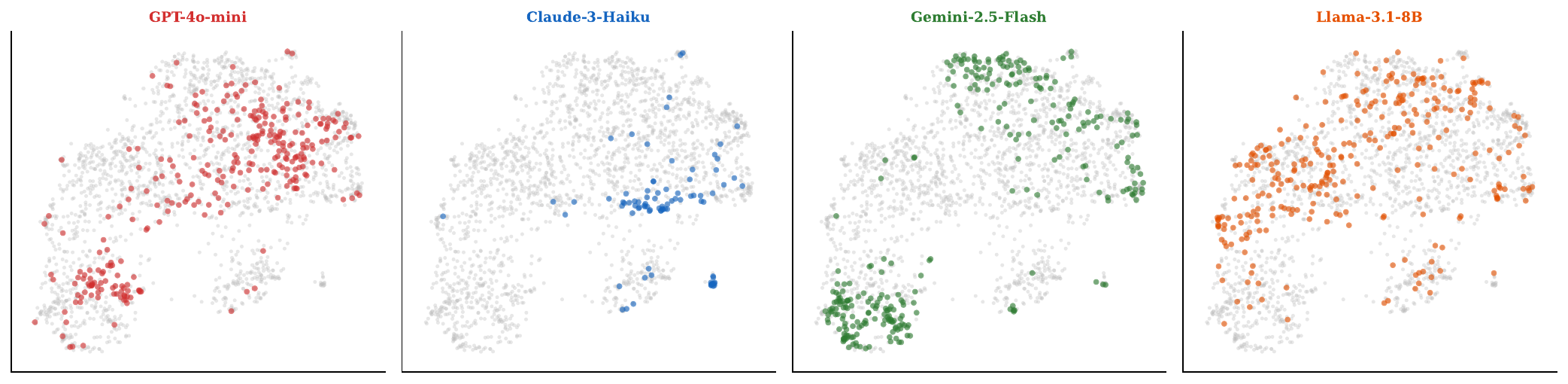}
\caption{Faceted UMAP projection of Sentence-BERT embeddings for outputs targeting Muslims. Four representative generators are shown; each panel highlights one model's outputs (colored) against the full eight-model distribution (gray).
}
\label{fig:umap_cross_model}
\end{figure*}

\mypara{Embedding-Space Separation}
To visualize cross-model complementarity, we fix the target identity group to Muslims and project the generated outputs from all eight models into a shared UMAP space using Sentence-BERT embeddings.
Fixing the target group controls for topical variation and allows us to focus on model-specific differences in style and framing. 
As shown in Figure~\ref{fig:umap_cross_model}, outputs from different models partially overlap but also occupy distinct subregions of the embedding space. 
API-gated models and open-weight models form visibly different regions, and individual models also show separation within each group. 
This suggests that different generators share a common semantic core while still introducing model-specific stylistic and rhetorical variation.

\mypara{Complementarity of Generators}
To further quantify each model's contribution, we compute its \emph{marginal coverage}: the fraction of the combined embedding space that is uniquely closest to samples from that model.
Let $\mathcal{X}_m$ denote the set of embedded samples generated by model $m$, and let $\mathcal{A}$ be a shared set of anchor points representing the combined embedding space, obtained by clustering or uniformly subsampling $\bigcup_m \mathcal{X}_m$. 
Each anchor point is assigned to the generator whose samples are closest to it:
\[
a(u)=\arg\min_{m\in\{1,\ldots,M\}} d(u,\mathcal{X}_m)
\]
\[
d(u,\mathcal{X}_m)=\min_{z\in\mathcal{X}_m}\|u-z\|_2 ,
\quad u\in\mathcal{A}.
\]
The marginal coverage of generator $m$ is defined as
\[
\mathrm{MC}(m)=
\frac{|\{u\in\mathcal{A}:a(u)=m\}|}{|\mathcal{A}|}.
\]
This metric measures the fraction of the shared embedding space that is uniquely closest to samples from generator $m$, rather than merely counting how many samples the generator produces.

As shown in Table~\ref{tab:marginal_coverage}, no single model dominates the output space. 
The most prolific model, Llama-3-8B, produces 154K usable samples but accounts for only 22.4\% of unique coverage. 
Conversely, Claude produces only 19K usable samples but still contributes 5.8\% of unique coverage. 
Removing any single model changes the overall fill distance by 3--8\%, confirming that each generator contributes non-redundant distributional mass.

\begin{table}[t!]
\centering
\caption{Marginal coverage of each generator, measured as the fraction of embedding-space regions uniquely closest to that model's outputs.}
\label{tab:marginal_coverage}
\small
\setlength{\tabcolsep}{5pt}
\begin{tabular}{lcc}
\toprule
\textbf{Model} & \textbf{\#Samples} & \textbf{Coverage (\%)} \\
\midrule
Llama-3-8B      & 154{,}190 & 22.4 \\
Gemma-2-9B      & 139{,}326 & 18.1 \\
Ministral-8B    & 122{,}592 & 14.7 \\
DeepSeek-R1-8B  & 111{,}018 & 13.2 \\
GPT-4o-mini     & 109{,}586 & 11.5 \\
Qwen-2.5-7B     & 104{,}218 & 9.8 \\
Gemini          & 31{,}223  & 4.5 \\
Claude          & 19{,}130  & 5.8 \\
\bottomrule
\end{tabular}
\end{table}

\mypara{Summary}
These results support our multi-model design. 
Different generators contribute complementary outputs, reducing dependence on any single model and broadening the distributional coverage of \method.

\subsection{Scalability}
\label{sec:efficiency}

In addition to diversity, we investigate the scalability of our \method by reporting the throughput in the generation and annotation processes.

\mypara{Generation Throughput}
Open-weight models served via vLLM on a single A100 GPU process 155{,}380 prompts in 3--9 hours, while API-gated models require approximately 5--23 hours depending on rate limits.
Across all eight generators, the total generation cost is approximately \$120 in API fees plus around 40 GPU-hours for open-weight models.
Detailed per-model efficiency statistics are reported in Appendix~\ref{app:generation_efficiency}.

\mypara{Annotation Throughput}
Our three-judge voting procedure is also parallelizable. 
With 16-thread parallelism and all judges queried in parallel, annotation processes approximately 10{,}000 samples per hour per model. 
Annotating the full 791K-sample corpus takes approximately 80 hours of wall-clock time when generation sources are processed in parallel. 
The annotation cost is dominated by API fees for the judge models.

\mypara{Summary}
Our pipeline is horizontally scalable at each stage. 
Persona selection via FPS is a one-time computation and runs in about 30 seconds for our persona pool. 
Generation is parallel across personas, identity groups, and generators, while annotation is parallel across samples. 
Adding a new generator only requires running generation and annotation for that model; the persona pool, selection procedure, and prompt design are shared. 
This modularity allows \method to be incrementally expanded as new models, identity groups, or hate targets emerge.

\subsection{Human Evaluation}
\label{sec:human_eval}

We further conduct human evaluations to assess three aspects of \method:
(1) whether the labels produced by our multi-judge annotation pipeline align with human judgments,
(2) whether the extracted personas faithfully represent their source posts without introducing excessive stereotypical characterization,
and (3) whether the generated samples resemble naturally occurring human-written hate speech.

Following standard sample-size estimation for proportions~\cite{C77}, approximately 385 samples are required to achieve a $\pm0.05$ margin of error at the 95\% confidence level under the conservative assumption $p=0.5$.
We therefore use 400 samples for each evaluation, with each sample independently evaluated by three human annotators.
Detailed annotation procedures and results are provided in Appendix~\ref{app:human_eval}.

\mypara{Annotation Reliability}
The human majority agrees with the six-judge labels at 87.0\% accuracy, 89.3\% F1, and Cohen's $\kappa=0.727$, compared with an average pairwise human--human agreement of $\kappa=0.411$.
On the 281 samples where all annotators agree, \method matches the unanimous human label in 276 cases (98.2\%), indicating that the six-judge labels align closely with human consensus, especially on clear cases.

\mypara{Persona Quality}
The extracted personas receive an average fidelity score of 4.2/5, with 85.4\% of ratings at least 4/5, while only 3.2\% are flagged as stereotypical.
Overall, the extracted personas largely reflect the characteristics expressed in their source posts, with little stereotypical characterization.

\mypara{Generated-Text Naturalness}
59.0\% of \method samples are classified as human-written, while the generated samples receive a high average naturalness score of 4.06/5, compared with 4.47/5 for human-written HateXplain samples.
These results suggest that a substantial fraction of \method-generated texts closely resemble naturally occurring human-written content.

\section{Enhancing Hate Speech Detection}
\label{sec:enhancement}

\subsection{Experimental Setup}
\label{sec:experimental_setup}

\subsubsection{Base Models and Training Settings}
\label{sec:base_models}

We train three types of detection models based on our \method, organized by how they are trained or adapted for hate speech detection.

\mypara{Encoders to Train From Scratch}
We first fine-tune four general-purpose transformer encoders that have not been specifically trained for hate speech detection. 
Each model is trained with a binary classification head on top of the \texttt{[CLS]} representation: \textbf{BERT-base} (110M parameters)~\cite{DCLT19}, \textbf{RoBERTa-base} (125M)~\cite{LOGDJCLLZS19}, \textbf{DeBERTa-v3} (184M)~\cite{HGC23}, and \textbf{XLM-RoBERTa-base} (278M)~\cite{CKGCWGGOZS20}. 
This setting tests whether \method provides sufficient supervision to train effective hate speech detectors from general-purpose language encoders.

\mypara{Pretrained Encoders to Fine-Tune}
We also evaluate whether \method provides complementary signal for models that have already been trained on hate speech data. 
For this setting, we continue fine-tuning three pretrained detectors on \method: \textbf{Detoxify-BERT} and \textbf{Detoxify-RoBERTa}~\cite{Detoxify}, corresponding to the ``original'' BERT-based and ``unbiased'' RoBERTa-based variants of Detoxify, and \textbf{LFTW-RoBERTa}~\cite{VTWK21}, a RoBERTa model trained through four rounds of dynamically collected hate speech data in the DynaBench setting. 
This setting examines whether persona-driven synthetic data improves detectors beyond existing hate speech training corpora.

\mypara{LLMs to Fine-Tune}
Finally, we collect four LLMs and fine-tune them on \method via LoRA.
Specifically, we adopt \textbf{Qwen-2.5-0.5B}~\cite{T243}, \textbf{Llama-3.2-1B}, \textbf{Llama-3.2-3B}~\cite{llama3_2_blog}, and \textbf{Llama-3.1-8B}~\cite{llama3_1_blog}, from Qwen and Meta, respectively.
These models are trained to perform binary hate speech classification in a generative format, outputting either ``yes'' or ``no''. 
This setting tests whether \method can train compact generative classifiers and whether model scale affects the benefit of synthetic training data.

\subsubsection{Baseline Methods}

To compare our trained detectors, we consider 14 representative baselines spanning pretrained encoders, commercial LLM APIs, large open-source LLMs, and moderation tools.

\mypara{Pretrained Encoders}
We consider three pretrained encoder-based detectors.
Detoxify is a Transformer-based toxicity classifier trained on the Jigsaw Toxic Comment benchmarks~\cite{Detoxify}. 
LFTW is a RoBERTa-based hate speech detector trained on the dynamically generated dataset from \emph{Learning from the Worst}~\cite{VTWK21}. 
Cardiff-RoBERTa is a Twitter-RoBERTa model fine-tuned for binary hate speech classification using 13 English hate speech datasets~\cite{cardiffnlp_hate_latest}.

\mypara{LLM APIs}
We include six commercial LLM APIs as prompt-based baselines for hate speech classification:
\textbf{GPT-4o-mini}~\cite{GPT-4o-mini},
\textbf{Claude-Sonnet-4}~\cite{Claude4},
\textbf{Gemini-2.5-Flash}~\cite{gemini_flash},
\textbf{Claude Opus 4.6}~\cite{claude_opus_46}, \textbf{Gem- ini-2.5-Pro}~\cite{gemini_pro}, and \textbf{DeepSeek-V4-Flash}~\cite{deepseek_v4_flash}.

\mypara{Large Open-Source LLMs}
We further evaluate large open-source LLMs in the zero-shot setting, including \textbf{Llama-3.1-70B}~\cite{llama3_1_blog}, \textbf{Qwen 3-32B}~\cite{qwen3}, and \textbf{Mistral-Small-24B}~\cite{mistral_small_3}.
These models provide stronger scaling baselines for comparing compact detectors trained on \method with much larger general-purpose LLMs.

\mypara{Moderation Tools}
We also include two moderation-oriented tools as category-based baselines.
Llama Guard 3 is a Llama-based safety classifier from Meta that classifies prompts and responses under a safety taxonomy; we use its \texttt{S10: Hate} category as the hate prediction~\cite{Llama_Guard_3}.
OpenAI Moderation is a hosted moderation endpoint for detecting harmful content; we use its \texttt{hate} categories as hate predictions~\cite{OpenAI_Moderation}.

\subsubsection{Evaluation Datasets}

We evaluate on four widely used hate speech detection benchmarks. 
Although Section~\ref{sec:dataset_limitations} discusses limitations of these datasets, they remain well-suited for evaluation because they are established benchmarks and cover different data sources and annotation settings. 
This allows us to test whether models trained on \method generalize beyond their synthetic training distribution.

\mypara{HateBenchSet} (7{,}838 samples) 
An LLM-generated benchmark covering 34 identity subgroups.

\mypara{HateXplain} (20{,}148 samples) A human-authored Twitter/Gab benchmark with target-group annotations.

\mypara{MHS} (39{,}565 samples) A social media benchmark with severity-based annotations from diverse annotators.

\mypara{Davidson} (5{,}593 samples after filtering) A Twitter benchmark used to test transfer under a different hate/offensive label taxonomy.

For multi-class datasets, we focus on binary hate speech detection by retaining hate and non-hate samples and removing samples labeled as offensive.

\subsubsection{Metrics}

We report accuracy and F1 for the binary classification task. 
For group-level evaluation, we report per-group F1, macro-F1, and the \emph{equity gap}, i.e., the difference between the best- and worst-performing groups. 
We use F1 as the primary metric.

\subsection{Performance of Training From Scratch on \method}
\label{sec:train_from_scratch}

We first evaluate \method as a training resource for hate speech detection, focusing on encoders trained on \method from scratch.
Table~\ref{tab:scratch_results} compares these models with pretrained detectors, commercial APIs, moderation tools, {and large open-source LLMs}.

Among the pretrained detectors, Cardiff-RoBERTa achieves 78.5\% average F1.
The commercial APIs achieve 77.1--80.8\% average F1.
Among the large open-source LLMs, Llama-3.1-70B performs best with 81.1\% average F1, followed by Qwen3-32B at 78.9\%.

Although the encoders trained on \method have no prior hate-speech-specific training, BERT, RoBERTa, and DeBERTa-v3 achieve {73.4\%, 74.6\%, and 77.5\%} average F1 across the four benchmarks, respectively (averaged over five random seeds).

The 184M-parameter DeBERTa-v3 is only 1.1 F1 points below Claude Opus 4.6 and 1.4 points below Qwen3-32B, while offering lower local deployment costs and faster inference.
On a single A100 GPU, it processes 520 samples/s, about 11$\times$ faster than Qwen3-32B and 96$\times$ faster than Claude Opus 4.6 under our evaluation setup.
Detailed efficiency and cost results are provided in Appendix~\ref{app:inference_efficiency}.

Notably, Llama-3.1-70B is the only stronger baseline that outperforms DeBERTa-v3 on all four benchmarks.

Overall, \method provides effective supervision for training hate speech detectors without relying on human-authored hate speech training data.

\begin{table*}[t!]
\centering
\caption{Training from scratch on \method. We report per-dataset F1 scores (\%) and average accuracy, F1, and macro-F1 across four benchmarks. Commercial APIs and pretrained detectors are included as reference baselines.}
\label{tab:scratch_results}
\small
\setlength{\tabcolsep}{4.5pt}
\begin{tabular}{llcccc|ccc}
\toprule
\textbf{Group} & \textbf{Model} & \textbf{HX} & \textbf{HB} & \textbf{DV} & \textbf{MHS} & \textbf{Acc} & \textbf{F1} & \textbf{M-F1} \\
\midrule
\multirow{4}{*}{Pretrained Encoders} 
 & Detoxify-BERT         & 63.5 & 75.4 & 73.6 & 71.8 & 73.9 & 71.1 & 73.1 \\
 & Detoxify-RoBERTa      & 63.7 & 78.1 & 71.9 & 71.8 & 74.4 & 71.4 & 73.6 \\
 & LFTW-RoBERTa          & 72.4 & 79.4 & 82.3 & 66.7 & 77.1 & 75.2 & 76.4 \\
 & Cardiff-RoBERTa       & 71.7 & 82.7 & 90.8 & 68.8 & 83.1 & 78.5 & 82.1 \\
\midrule
\multirow{6}{*}{LLM APIs}
 & GPT-4o-mini           & 79.3 & 89.2 & 80.4 & 70.1 & 81.9 & 79.7 & 81.0 \\
 & Claude-4-Sonnet       & 82.8 & 88.8 & 81.5 & 68.8 & 83.7 & 80.5 & 82.8 \\
 & Gemini-2.5-Flash      & 84.4 & 87.8 & 82.5 & 68.5 & 84.5 & 80.8 & 83.5 \\
 & {Claude Opus 4.6}     
 & {81.6} 
 & {89.0} 
 & {78.1} 
 & {65.8} 
 & {82.8} 
 & {78.6} 
 & {81.7} \\
 
 & {Gemini-2.5-Pro}      
 & {87.5} 
 & {87.1} 
 & {69.9} 
 & {48.0} 
 & {83.1} 
 & {73.1} 
 & {80.1} \\

 & {DeepSeek V4-Flash}   
 & {83.6} 
 & {88.4} 
 & {77.8} 
 & {62.1} 
 & {83.5} 
 & {78.0} 
 & {82.1} \\
\midrule
\multirow{2}{*}{Moderation Tools}
 & Llama Guard 3         & 79.7 & 85.5 & 76.0 & 56.2 & 81.4 & 74.3 & 79.5 \\
 & OpenAI Moderation     & 75.6 & 84.3 & 76.8 & 57.8 & 80.4 & 73.6 & 78.6 \\
\midrule
\multirow{3}{*}{{Open-source LLMs}}
 & {Llama-3.1-70B}
 & {79.1}
 & {89.1}
 & {83.9}
 & {72.3}
 & {83.6}
 & {81.1}
 & {82.9} \\

 & {Qwen3-32B}
 & {75.4}
 & {87.8}
 & {80.2}
 & {72.2}
 & {80.6}
 & {78.9}
 & {79.9} \\

 & {Mistral-Small-24B}
 & {55.2}
 & {61.0}
 & {72.3}
 & {60.9}
 & {73.6}
 & {62.3}
 & {70.8} \\
\midrule
\multirow{4}{*}{Encoders + \method}
 & BERT                  & 64.5 & 86.9 & 74.3 & 67.8 & 77.0 & 73.4 & 75.9 \\
 & RoBERTa               & 66.7 & 87.1 & 74.7 & 69.9 & 76.3 & 74.6 & 75.4 \\
 & DeBERTa-v3            & 71.3 & 88.9 & 78.8 & 70.9 & 79.4 & 77.5 & 78.5 \\
 & XLM-RoBERTa           & 57.3 & 86.8 & 59.7 & 61.8 & 73.2 & 66.4 & 71.2 \\
\bottomrule
\end{tabular}
\end{table*}

\subsection{Performance of Fine-Tuning on \method}
\label{sec:finetune_pretrained}

We next examine whether \method provides a complementary training signal for models that already have hate speech or safety-related training, i.e., whether fine-tuning on \method can enhance detection performance.
Table~\ref{tab:finetune_results} reports paired results before and after fine-tuning each model on \method{, with the fine-tuned results averaged over five random seeds and statistical significance tested against the corresponding baseline}.

We observe that fine-tuning on \method {significantly improves all four pretrained encoder models}.
LFTW-RoBERTa improves from 75.2\% to {77.3\%} average F1, while Detoxify-BERT improves from 71.1\% to {75.4\%}, and Detoxify-RoBERTa improves from 71.4\% to {74.2\%}.
Cardiff-RoBERTa, the strongest pretrained encoder baseline, also improves from 78.5\% to 79.0\% average F1, with statistically significant gains observed for all four encoders.

The LLMs also benefit substantially from supervised fine-tuning on \method.
Qwen2.5-0.5B improves from 9.1\% to 67.4\% average F1.
Llama-3.2-1B improves from 34.2\% to 73.1\% average F1, and Llama-3.2-3B improves from 47.9\% to 78.7\%.
The largest model, Llama-3.1-8B, already has stronger zero-shot performance (70.7\%) but still improves to 78.4\%, corresponding to a 7.7-point gain.
These improvements are statistically significant for all four LLMs ($p<0.001$).

Overall, the consistent improvements across all pretrained encoders and LLMs suggest that \method is not redundant with existing hate speech or safety-related training data.
Instead, its persona-driven and LLM-generated content provides complementary supervision that improves performance on both generated and human-authored hate speech benchmarks.

\begin{table*}[t!]
\centering
\caption{Fine-tuning pretrained models with \method.
We report F1 scores (\%) across four benchmarks before and after \method fine-tuning.
$\Delta$ denotes the change in average F1.
{\textit{+\method} results are averaged over 5 random seeds; significance of improvement over zero-shot is tested via one-tailed one-sample $t$-test ($^{*}p<0.05$, $^{**}p<0.01$, $^{***}p<0.001$).}}
\label{tab:finetune_results}
\small
\setlength{\tabcolsep}{5pt}
\begin{tabular}{llcccc|c|r}
\toprule
\textbf{Model} & \textbf{Setting} & \textbf{HX} & \textbf{HB} & \textbf{DV} & \textbf{MHS} & \textbf{Avg} & \textbf{$\Delta$} \\
\midrule

Detoxify-BERT    & zero-shot & 63.5 & 75.4 & 73.6 & 71.8 & 71.1 & -- \\
                 & +\method  & 65.0 & 86.8 & 79.0 & 70.8 & 75.4 & $+4.3^{***}$ \\

\midrule
Detoxify-RoBERTa & zero-shot & 63.7 & 78.1 & 71.9 & 71.8 & 71.4 & -- \\
                 & +\method  & 65.7 & 87.1 & 74.6 & 69.2 & 74.2 & $+2.7^{***}$ \\

\midrule
LFTW-RoBERTa     & zero-shot & 72.4 & 79.4 & 82.3 & 66.7 & 75.2 & -- \\
                 & +\method  & 69.2 & 87.8 & 81.3 & 70.8 & 77.3 & $+2.1^{**}$ \\

\midrule
Cardiff-RoBERTa  & zero-shot & 71.7 & 82.7 & 90.8 & 68.8 & 78.5 & -- \\
                 & +\method  & 74.1 & 87.1 & 84.6 & 70.1 & 79.0 & $+0.5^{**}$ \\

\midrule
Qwen-2.5-0.5B    & zero-shot & 1.8 & 24.6 & 5.4 & 4.7 & 9.1 & -- \\
                 & +\method  & 59.7 & 83.4 & 62.4 & 64.1 & 67.4 & $+58.3^{***}$ \\

\midrule
Llama-3.2-1B     & zero-shot & 32.6 & 43.9 & 23.0 & 37.1 & 34.2 & -- \\
                 & +\method  & 67.5 & 83.4 & 72.7 & 68.9 & 73.1 & $+38.9^{***}$ \\

\midrule
Llama-3.2-3B     & zero-shot & 25.8 & 79.5 & 42.0 & 44.2 & 47.9 & -- \\
                 & +\method  & 77.1 & 86.1 & 81.8 & 70.0 & 78.7 & $+30.8^{***}$ \\

\midrule
Llama-3.1-8B     & zero-shot & 69.3 & 78.3 & 74.4 & 60.8 & 70.7 & -- \\
                 & +\method  & 77.1 & 85.8 & 82.6 & 68.0 & 78.4 & $+7.7^{***}$ \\

\bottomrule
\end{tabular}
\end{table*}

\subsection{Worst-Group Improvement}
\label{sec:worst_group}

As shown in Section~\ref{sec:limitation_detectors}, one limitation of hate speech detectors is that the detection performance varies substantially across identity groups.
A natural mitigation strategy is to collect or synthesize more data for the weakest groups and train models specifically on those groups. 
We test whether such targeted identity group training is sufficient, and compare it with training on the full, diverse \method dataset.

We first identify 12 low-performing groups under zero-shot Qwen-2.5-0.5B on HateBenchSet and train separate single-group models using approximately 2K samples from each target group extracted from \method. 
For each group, we compare three settings: zero-shot inference, targeted single-group training, and full multi-group training using approximately 67K samples from all 34 groups (\method). 
All models use the same Qwen-2.5-0.5B architecture with LoRA fine-tuning.

As shown in Table~\ref{tab:targeted_training}, targeted training substantially improves performance over the zero-shot baseline for every group. 
The gains range from +2.6 percentage points for Men to +47.8 points for Immigrants, with an average improvement of +27.5 points. 
Groups with particularly low zero-shot performance, such as Physical Disability and Gay, benefit strongly from targeted training, suggesting that additional group-specific supervision can help address severe under-detection.

However, targeted training is consistently outperformed by full multi-group \method training. 
The full model achieves higher F1 than the corresponding single-group model for all 12 target groups, with an average F1 of 79.9\% compared with 55.6\% for targeted training. 
This suggests that broad group coverage provides useful cross-group transfer: patterns learned from diverse hate targets help improve detection even for the weakest groups, beyond what can be achieved by concentrating training data on the target group alone.

\begin{table}[t!]
\centering
\caption{F1 scores (\%) on HateBenchSet for low-performing target groups under three training settings using Qwen-2.5-0.5B.
Targeted training uses approximately 2K samples from the target group only; Full training uses approximately 67K samples from all 34 groups (\method).}
\label{tab:targeted_training}
\small
\setlength{\tabcolsep}{5pt}
\begin{tabular}{lccc|cc}
\toprule
\textbf{Target Group} & \textbf{Zero-shot} & \textbf{Targeted} & \textbf{Full} & \textbf{$\Delta_{\text{tgt}}$} & \textbf{$\Delta_{\text{full}}$} \\
\midrule
Immigrants          & 25.5 & 73.3 & 87.9 & +47.8 & +62.4 \\
Phys.\ Disability   &  9.8 & 57.0 & 75.5 & +47.2 & +65.7 \\
Migrant Workers     & 33.2 & 69.2 & 84.3 & +36.1 & +51.1 \\
Muslims             & 37.9 & 71.3 & 86.6 & +33.4 & +48.7 \\
Asian               & 29.2 & 62.5 & 67.1 & +33.3 & +37.9 \\
Women               & 23.3 & 55.2 & 79.6 & +31.9 & +56.3 \\
Gay                 & 18.6 & 47.7 & 68.0 & +29.2 & +49.4 \\
Cogn.\ Disability   & 17.6 & 42.7 & 77.6 & +25.1 & +60.0 \\
Black               & 39.3 & 63.4 & 79.6 & +24.1 & +40.3 \\
White               & 30.4 & 45.7 & 81.7 & +15.3 & +51.3 \\
Christians          & 33.7 & 37.7 & 86.5 &  +4.1 & +52.8 \\
Men                 & 38.5 & 41.1 & 84.5 &  +2.6 & +46.0 \\
\midrule
\textbf{Average}    & 28.1 & 55.6 & 79.9 & +27.5 & +51.8 \\
\bottomrule
\end{tabular}
\end{table}

\subsection{Cross-Dataset Generalization}
\label{sec:cross_dataset_results}

Another common failure mode of hate speech detectors is strong performance on one dataset but poor transfer to others, as illustrated in Section~\ref{sec:limitation_detectors}. 
We therefore examine whether \method-trained models remain effective across benchmarks with different sources, annotation schemes, and label definitions.

\mypara{Consistent Performance Across Benchmarks}
\method-trained encoders show relatively stable performance across the four evaluation datasets. 
For example, DeBERTa-v3 trained on \method achieves 70.9\% F1 on HateXplain and 89.8\% on HateBenchSet (see Table~\ref{tab:scratch_results}), while LFTW-RoBERTa+\method ranges from 69.4\% to 88.7\% (see Table~\ref{tab:finetune_results}).
We find that all detectors trained or fine-tuned on \method achieve good performance on LLM-generated HateBenchSet, and maintain competitive results on human-authored datasets such as HateXplain, Davidson, and MHS.
We also see that our \method could help enhance the hate speech detection performance in most cases.
This is notable because \method is fully synthetic, whereas the external benchmarks contain human-authored content from different platforms. 
The results suggest that our persona-driven synthetic data captures generalizable hate speech patterns, including group targeting, derogatory framing, and recurring rhetorical structures, rather than only model-specific artifacts.

\mypara{Robustness Across Annotation Schemes}
Note that the evaluation benchmarks use different labeling schemes: HateBenchSet and Davidson distinguish hate from offensive language, HateXplain uses hate/offensive/normal labels, and MHS uses severity-based annotations. 
Despite these differences, the best \method-trained models achieve F1 above 70\% on three of four benchmarks. 
This suggests that the binary labels produced by our annotation pipeline align reasonably well with multiple human annotation standards.

\subsection{Adaptation to Evolving Hate Distributions}
\label{sec:temporal_generalization}

A challenge for hate speech detection is whether it can adapt to \emph{newly emerging} hate, i.e., content involving topics, events, or rhetoric that emerges after existing datasets were collected.
We study this setting using the New-Wave Hate benchmark~\cite{VGROCZH24}, which covers six emerging hate topics: anti-Asian hate (Asian), ageism, vaccine-related hate (Vaccine), mask-related hate (Mask), Capitol riot rhetoric (Capitol), and Russia--Ukraine conflict rhetoric (Rus--Ukr).
We quickly adapt our \method accordingly: we first sample 200 personas from the persona pool; then we prompt the generators conditioned on the persona, sentiment, and the hate topic (replacing the target identity group); finally, we adopt the multi-judge annotation for labeling and filtering.
Overall, we obtain 540 balanced labeled samples for these emerging hate topics.

Based on these labeled samples, we fine-tune two pretrained encoders, LFTW-RoBERTa and Cardiff-RoBERTa, as well as Llama-3.1-8B, to evaluate how \method can adapt different detector architectures to emerging hate topics.
Table~\ref{tab:newwave_results} compares these adapted models with commercial APIs, moderation tools, zero-shot open-source LLMs, and pretrained encoders.

Among the commercial APIs and moderation tools, GPT-4o-mini achieves the highest overall F1 of 69.8\%, but its performance varies considerably across categories, with a 46.5-point gap between its best and worst results.
Among the zero-shot open-source LLMs, Qwen3-32B reaches 68.3\% overall F1 with a 37.6-point gap, while Llama-3.1-8B reaches 61.0\% with a smaller 17.2-point gap.
The pretrained encoders achieve 63.2\% overall F1 for LFTW-RoBERTa and 61.1\% for Cardiff-RoBERTa.

To understand the uneven performance of the API baselines, we further analyze their prediction outputs.
Invalid or unparsable responses are negligible (e.g., 0.025\% for Gemini-2.5-Flash and 0.1\% for Claude Opus 4.6), indicating that the low scores are not primarily caused by failed API responses.
Instead, several APIs show low recall on emerging categories.
For example, Gemini-2.5-Flash achieves only 3\% recall on Ageism and Mask, 5\% on Vaccine, and 1\% on Capitol, despite returning valid predictions for almost all samples.
This suggests that the uneven performance mainly reflects category-dependent under-detection rather than refusals or parsing failures.

Fine-tuning on \method improves all three adapted models.
LFTW-RoBERTa increases from 63.2\% to 69.9\% overall F1, while Cardiff-RoBERTa increases from 61.1\% to 72.7\%.
Their cross-category gaps also decrease from 35.0 to 22.3 points and from 36.0 to 18.0 points, respectively.
Llama-3.1-8B improves from 61.0\% to 76.7\% overall F1, with its cross-category gap decreasing from 17.2 to 15.3 points.

Llama-3.1-8B + \method achieves the highest overall F1 and the smallest cross-category gap among all evaluated models.
It outperforms all commercial APIs on four of the six categories, with the largest margin on mask-related hate (+31.4 points over GPT-4o-mini).

These results suggest that scalable persona-driven synthesis can provide an effective and efficient mechanism for updating detectors as new hate targets and rhetoric emerge.

\begin{table*}[t!]
\centering
\caption{Per-category F1 scores (\%) on the New-Wave Hate benchmark.
{\method adapts pretrained encoders and Llama-3.1-8B to emerging hate topics using 540 balanced samples generated and annotated by our \method.}}
\label{tab:newwave_results}
\small
\setlength{\tabcolsep}{4pt}
\begin{tabular}{lcccccc|c|c}
\toprule
\textbf{Model} & \textbf{Asian} & \textbf{Ageism} & \textbf{Mask} & \textbf{Vaccine} & \textbf{Capitol} & \textbf{Rus--Ukr} & \textbf{Gap} & \textbf{Overall} \\
\midrule

\multicolumn{9}{l}{\textit{Commercial APIs \& Moderation Tools}} \\

OpenAI Moderation   
& 65.9 & 42.0 & 12.4 & 31.1 & 24.3 & 67.6 & 55.2 & 48.7 \\

Claude-4-Sonnet     
& 83.3 & 27.5 & 21.7 & 27.9 & 20.3 & 65.9 & 63.0 & 55.5 \\

{Claude Opus 4.6}
& 81.9 & 40.8 & 29.8 & 34.5 & 27.1 & 71.6 & 54.8 & 58.8 \\

Gemini-2.5-Flash    
& 77.6 & 34.7 & 28.2 & 42.3 & 41.4 & 64.3 & 49.4 & 57.5 \\

GPT-4o-mini         
& \textbf{85.5} & 52.1 & 39.0 & 54.7 & 57.9 & \textbf{81.5} & 46.5 & 69.8 \\

DeepSeek V4-Flash
& 82.8 & 10.4 & 14.4 & 16.4 & 2.5 & 74.0 & 80.3 & 53.1 \\

\midrule
\multicolumn{9}{l}{{\textit{Open-source LLMs (zero-shot)}}} \\

Mistral-Small-24B
& 62.4 & 25.0 & 21.4 & 42.4 & 27.2 & 52.3 & 41.0 & 45.2 \\

Qwen3-32B
& 85.0 & 49.5 & 47.8 & 58.7 & 47.4 & 77.7 & 37.6 & 68.3 \\

Llama-3.1-8B zero-shot 
& 67.0 & 49.8 & 58.4 & 59.9 & 55.0 & 65.2 & 17.2 & 61.0 \\

\midrule
\multicolumn{9}{l}{{\textit{Pretrained Encoders}}} \\

LFTW-RoBERTa
& 75.0 & 40.0 & 57.1 & 51.2 & 54.7 & 71.8 & 35.0 & 63.2 \\

Cardiff-RoBERTa
& 76.7 & 42.5 & 40.7 & 56.7 & 42.0 & 60.1 & 36.0 & 61.1 \\

\midrule
\multicolumn{9}{l}{{\textit{Fine-tuned on \method}}} \\

LFTW-RoBERTa + \method
& 76.7 & 54.4 & 70.2 & 63.1 & 64.9 & 75.7 & 22.3 & 69.9 \\

Cardiff-RoBERTa + \method
& 79.5 & 61.5 & 67.9 & 71.8 & 65.0 & 76.7 & 18.0 & 72.7 \\

Llama-3.1-8B + \method
& 83.8
& \textbf{68.5}
& \textbf{70.4}
& \textbf{73.0}
& \textbf{73.7}
& 77.9
& \textbf{15.3}
& \textbf{76.7} \\

\bottomrule
\end{tabular}
\end{table*}

\subsection{Ablation Study: Does Diversity Matter Beyond Scale?}
\label{sec:ablation_diversity}

We conduct ablation studies to examine whether the gains from \method come only from increasing the number of training samples, or from increasing the diversity of the training data.
This distinction is important because \method is designed not merely to scale up synthetic data, but to cover diverse identity groups, generator behaviors, and persona-conditioned expressions.
We focus on two diversity axes in this section: identity group coverage and generator diversity.
All experiments use DeBERTa-v3 with the same training procedure, and all sampled training sets are label-balanced.

\mypara{Size-Controlled Diversity}
We first control the number of training samples and vary only the diversity of the data.
Table~\ref{tab:diversity_ablation} reports results at two data scales.
At the 2K-sample scale, training on one group from one generator achieves 67.6\% average F1.
Keeping the same sample size but adding all eight generators improves the average F1 to 68.9\%.
Using all 34 identity groups and all eight generators further improves it to 69.3\%.
This shows that, even when the amount of training data is fixed, broader coverage over generators and identity groups improves cross-dataset generalization.

The effect becomes clearer at the 10K-sample scale.
The restricted setting with one generator and five groups achieves only 69.1\% average F1.
Expanding identity-group coverage while keeping one generator improves performance to 75.8\%, and expanding generator coverage while keeping five groups improves performance to 76.1\%.
Thus, neither data size nor a single diversity source is sufficient by itself.
Group diversity and generator diversity provide complementary benefits, and removing both substantially weakens the trained detector.

\mypara{Scaling Identity Group Coverage}
We next test whether identity-group diversity remains useful when the training set becomes larger.
Table~\ref{tab:diversity_large} compares models trained with 10, 17, and 34 identity groups while using all eight generators.
Training with 10 groups reaches 74.3\% average F1, expanding to 17 groups improves performance to 75.6\%, and using all 34 groups achieves 77.6\%.
The gain from 17 to 34 groups is larger than the gain from 10 to 17 groups, and the full 34-group setting remains the strongest among the group-scaling conditions.

These results indicate that identity-group coverage contributes to detector robustness beyond raw data volume.
A broader set of target groups exposes the detector to more varied hateful framings, stereotypes, and linguistic patterns, some of which transfer across groups and benchmarks.
This finding is consistent with the worst-group analysis in Section~\ref{sec:worst_group}, where full multi-group training outperforms targeted single-group training on weakly detected groups.

\mypara{Statistical Significance}
To verify that the observed diversity gains are robust to random variation, we repeat the key ablation experiments over five random seeds and conduct statistical significance tests on the average F1 across the four benchmarks.
Increasing generator diversity at the 10K scale yields a significant improvement of 3.9 F1 points ($p=0.004$), while increasing identity-group coverage yields a significant improvement of 10.2 points ($p=0.009$).
Combining both diversity dimensions also significantly outperforms the restricted single-group, single-generator setting, with an 8.1-point improvement ($p=0.011$).
These results confirm that the gains from broader generator and identity-group coverage are statistically robust.

Overall, the ablation confirms that \method's benefit comes from diversity-aware data construction, not simply from generating more synthetic samples.

\begin{table*}[t!]
\centering
\caption{Size-controlled diversity ablation using DeBERTa-v3.
All conditions are label-balanced.
Condition names use abbreviated settings: M denotes the number of generators (Mdl), and G denotes the number of identity groups (Grp).
The average F1 score is computed across four benchmarks.
Results averaged over 5 random seeds.
}
\label{tab:diversity_ablation}
\small
\setlength{\tabcolsep}{5pt}
\begin{tabular}{lccc|cccc|c}
\toprule
\textbf{Condition} & \textbf{$N$} & \textbf{Grp} & \textbf{Mdl} & \textbf{HX} & \textbf{HB} & \textbf{DV} & \textbf{MHS} & \textbf{Avg} \\
\midrule
\texttt{1M1G\_2K}   & 2K  &  1 & 1 & 63.7 & 83.6 & 48.5 & 62.2 & 64.5 \\
\texttt{8M1G\_2K}   & 2K  &  1 & 8 & 69.6 & 85.6 & 62.4 & 65.6 & 70.8 \\
\texttt{8M34G\_2K}  & 2K  & 34 & 8 & 67.8 & 87.1 & 66.9 & 68.4 & 72.6 \\
\midrule
\texttt{1M5G\_10K}  & 10K &  5 & 1 & 69.7 & 87.7 & 67.8 & 64.5 & 72.4 \\
\texttt{1M34G\_10K} & 10K & 34 & 1 & 69.7 & 88.4 & 72.8 & 67.9 & 74.7 \\
\texttt{8M5G\_10K}  & 10K &  5 & 8 & 71.7 & 88.6 & 76.1 & 69.1 & 76.4 \\
\bottomrule
\end{tabular}
\end{table*}

\begin{table*}[t!]
\centering
\caption{Effect of identity group coverage at larger training scales using DeBERTa-v3.
All conditions use eight generators.
Condition names use abbreviated settings: M denotes the number of generators (Mdl), and G denotes the number of identity groups (Grp).
The average F1 score is computed across four benchmarks.
}
\label{tab:diversity_large}
\small
\setlength{\tabcolsep}{5pt}
\begin{tabular}{lcc|cccc|c}
\toprule
\textbf{Condition} & \textbf{$N$} & \textbf{Grp} & \textbf{HX} & \textbf{HB} & \textbf{DV} & \textbf{MHS} & \textbf{Avg} \\
\midrule
\texttt{8M10G}  & 20K & 10 & 71.5 & 89.4 & 70.2 & 66.1 & 74.3 \\
\texttt{8M17G}  & 34K & 17 & 69.5 & 88.7 & 74.6 & 69.6 & 75.6 \\
\texttt{1M34G}      & 51K & 34 & 70.9 & 88.3 & 76.7 & 70.1 & 76.5 \\
\texttt{8M34G} (full)  & 67K & 34 & 70.9 & 89.8 & 78.8 & 71.0 & 77.6 \\
\bottomrule
\end{tabular}
\end{table*}

\begin{table*}[t!]
\centering
\caption{{Statistical significance of key diversity ablations over five random seeds. $\Delta$ denotes the improvement in average F1 across the four benchmarks.}}
\label{tab:diversity_significance}
\small
\begin{tabular}{l l c c}
\toprule
Comparison & Effect & $\Delta$ F1 & $p$-value \\
\midrule
1M5G $\rightarrow$ 8M5G (10K)
& Generator diversity
& +3.9
& 0.004 \\
1M1G $\rightarrow$ 1M34G
& Group coverage
& +10.2
& 0.009 \\
1M1G $\rightarrow$ 8M34G (2K)
& Joint diversity
& +8.1
& 0.011 \\
\bottomrule
\end{tabular}
\end{table*}

\section{Limitations}
\label{sec:limitations}

\mypara{Synthetic Data as Training Data}
\method is designed as a synthetic data pipeline, not as a replacement for human-authored hate speech benchmarks.
Human-authored datasets remain important because they capture naturally occurring discourse, platform-specific norms, social context, and annotation standards that are difficult to fully reproduce through synthetic generation~\cite{TLMHFOR25}.
For this reason, we use \method to train or fine-tune detectors and rely on external benchmarks for evaluation.
The strong cross-dataset results suggest that persona-driven synthetic data can provide useful supervision, but they should not be interpreted as evidence that synthetic data alone fully captures real-world hate speech.

\mypara{Label Quality and Ambiguous Cases}
\method uses multi-judge annotation to reduce dependence on any single classifier or LLM judge.
However, hate speech labeling remains inherently difficult, especially for implicit hate, counterspeech, reclaimed slurs, quotation, and context-dependent cases.
Our binary labeling scheme supports large-scale detector training, but it cannot capture all levels of severity, intent, target ambiguity, or pragmatic context.
Future work could combine multi-judge annotation with expert or community-informed review for ambiguous samples, and extend the labels to severity, target, rationale, or explanation annotations.

\mypara{Scope and Generalization}
Our current pipeline focuses on English text and 34 identity groups following established hate speech benchmarks.
This design supports controlled comparison with prior work, but it does not cover all languages, regions, cultural contexts, or intersectional identities.
Hate speech also appears in multimodal and conversational settings, where images, memes, dialogue history, and platform context can change the meaning of a message.
Extending \method to multilingual, multimodal, and conversation-aware generation is an important direction for future work.

\section{Conclusion}

We presented \method, a scalable persona-driven data synthesis pipeline for hate speech analysis.
\method constructs a large persona pool from complementary sources, selects diverse personas through embedding-based sampling, generates group-targeted speech across 34 identity groups using multiple LLMs, and applies multi-judge annotation to obtain reliable hate speech labels.
The resulting corpus contains 791K valid samples, from which we curate a 67K group-balanced training set.

Besides diversity and scalability analysis, our experiments further show that persona-driven synthetic data improves hate speech detection across datasets and identity groups.
Models trained or fine-tuned on \method achieve strong cross-dataset performance, improve weak-group detection, and benefit from broader coverage over identity groups, generators, and personas.
These results suggest that diversity-aware data synthesis is a practical way to improve the robustness and generalizability of hate speech detectors.
We hope this work encourages future research that improves hate speech detection through better data design, in addition to better model design.

\bibliographystyle{plain}
\bibliography{normal_generated_py3}

@inproceedings{DCLT19,
author = {Jacob Devlin and Ming{-}Wei Chang and Kenton Lee and Kristina Toutanova},
title = {{BERT: Pre-training of Deep Bidirectional Transformers for Language Understanding}},
booktitle = {{Conference of the North American Chapter of the Association for Computational Linguistics: Human Language Technologies (NAACL-HLT)}},
pages = {4171-4186},
publisher = {ACL},
year = {2019}
}

@inproceedings{PZCSB20,
author = {Antonis Papasavva and Savvas Zannettou and Emiliano De Cristofaro and Gianluca Stringhini and Jeremy Blackburn},
title = {{Raiders of the Lost Kek: 3.5 Years of Augmented 4chan Posts from the Politically Incorrect Board}},
booktitle = {{International Conference on Web and Social Media (ICWSM)}},
pages = {885-894},
publisher = {AAAI},
year = {2020}
}

@inproceedings{DWMW17,
author = {Thomas Davidson and Dana Warmsley and Michael W. Macy and Ingmar Weber},
title = {{Automated Hate Speech Detection and the Problem of Offensive Language}},
booktitle = {{International Conference on Web and Social Media (ICWSM)}},
pages = {512-515},
publisher = {AAAI},
year = {2017}
}

@inproceedings{MSYBGM21,
author = {Binny Mathew and Punyajoy Saha and Seid Muhie Yimam and Chris Biemann and Pawan Goyal and Animesh Mukherjee},
title = {{HateXplain: {A} Benchmark Dataset for Explainable Hate Speech Detection}},
booktitle = {{AAAI Conference on Artificial Intelligence (AAAI)}},
pages = {14867-14875},
publisher = {AAAI},
year = {2021}
}

@inproceedings{RVNWMP21,
author = {Paul R{\"{o}}ttger and Bertie Vidgen and Dong Nguyen and Zeerak Waseem and Helen Z. Margetts and Janet B. Pierrehumbert},
title = {{HateCheck: Functional Tests for Hate Speech Detection Models}},
booktitle = {{Annual Meeting of the Association for Computational Linguistics and International Joint Conference on Natural Language Processing (ACL/IJCNLP)}},
pages = {41-58},
publisher = {ACL},
year = {2021}
}

@inproceedings{SBBSVK22,
author = {Pratik S. Sachdeva and Renata Barreto and Geoff Bacon and Alexander Sahn and Claudia von Vacano and Chris J. Kennedy},
title = {{The Measuring Hate Speech Corpus: Leveraging Rasch Measurement Theory for Data Perspectivism}},
booktitle = {{Workshop on Perspectivist Approaches to NLPerspectives (NLPerspectives)}},
pages = {83-94},
publisher = {ELRA},
year = {2022}
}

@inproceedings{VTWK21,
author = {Bertie Vidgen and Tristan Thrush and Zeerak Waseem and Douwe Kiela},
title = {{Learning from the Worst: Dynamically Generated Datasets to Improve Online Hate Detection}},
booktitle = {{Annual Meeting of the Association for Computational Linguistics and International Joint Conference on Natural Language Processing (ACL/IJCNLP)}},
pages = {1667-1682},
publisher = {ACL},
year = {2021}
}

@inproceedings{CBMG21,
author = {Tommaso Caselli and Valerio Basile and Jelena Mitrovic and Michael Granitzer},
title = {{{H}ate{BERT}: Retraining {BERT} for Abusive Language Detection in {E}nglish}},
booktitle = {{Workshop on Online Abuse and Harms (WOAH)}},
pages = {17-25},
publisher = {ACL},
year = {2021}
}

@inproceedings{WAM21,
author = {Tomer Wullach and Amir Adler and Einat Minkov},
title = {{Fight Fire with Fire: Fine-tuning Hate Detectors using Large Samples of Generated Hate Speech}},
booktitle = {{Findings of the Association for Computational Linguistics: EMNLP (EMNLP Findings)}},
pages = {4699-4705},
publisher = {ACL},
year = {2021}
}

@inproceedings{VGROCZH24,
author = {Nishant Vishwamitra and Keyan Guo and Farhan Tajwar Romit and Isabelle Ondracek and Long Cheng and Ziming Zhao and Hongxin Hu},
title = {{Moderating New Waves of Online Hate with Chain-of-Thought Reasoning in Large Language Models}},
booktitle = {{IEEE Symposium on Security and Privacy (S\&P)}},
publisher = {IEEE},
year = {2024}
}

@inproceedings{SWQBZZ25,
author = {Xinyue Shen and Yixin Wu and Yiting Qu and Michael Backes and Savvas Zannettou and Yang Zhang},
title = {{HateBench: Benchmarking Hate Speech Detectors on LLM-Generated Content and Hate Campaigns}},
booktitle = {{USENIX Security Symposium (USENIX Security)}},
publisher = {USENIX},
year = {2025}
}

@inproceedings{LTTSGMV22,
author = {Alyssa Lees and Vinh Q. Tran and Yi Tay and Jeffrey Sorensen and Jai Gupta and Donald Metzler and Lucy Vasserman},
title = {{A New Generation of Perspective API: Efficient Multilingual Character-level Transformers}},
booktitle = {{ACM Conference on Knowledge Discovery and Data Mining (KDD)}},
pages = {3197-3207},
publisher = {ACM},
year = {2022}
}

@inproceedings{ASK20,
author = {Pedro Alonso and Rajkumar Saini and Gy{\"{o}}rgy Kov{\'{a}}cs},
title = {{TheNorth at SemEval-2020 Task 12: Hate Speech Detection Using RoBERTa}},
booktitle = {{International Workshop on Semantic Evaluation (SemEval)}},
pages = {2197-2202},
publisher = {International Committee for Computational Linguistics},
year = {2020}
}

@inproceedings{BGGV17,
author = {Pinkesh Badjatiya and Shashank Gupta and Manish Gupta and Vasudeva Varma},
title = {{Deep Learning for Hate Speech Detection in Tweets}},
booktitle = {{International Conference on World Wide Web (WWW)}},
pages = {759-760},
publisher = {ACM},
year = {2017}
}

@inproceedings{BGMS21,
author = {Emily M. Bender and Timnit Gebru and Angelina McMillan{-}Major and Shmargaret Shmitchell},
title = {{On the Dangers of Stochastic Parrots: Can Language Models Be Too Big?}},
booktitle = {{Conference on Fairness, Accountability, and Transparency (FAccT)}},
pages = {610-623},
publisher = {ACM},
year = {2021}
}

@inproceedings{CKGCWGGOZS20,
author = {Alexis Conneau and Kartikay Khandelwal and Naman Goyal and Vishrav Chaudhary and Guillaume Wenzek and Francisco Guzm{\'{a}}n and Edouard Grave and Myle Ott and Luke Zettlemoyer and Veselin Stoyanov},
title = {{Unsupervised Cross-lingual Representation Learning at Scale}},
booktitle = {{Annual Meeting of the Association for Computational Linguistics (ACL)}},
pages = {8440-8451},
publisher = {ACL},
year = {2020}
}

@inproceedings{DRSC26,
author = {Saloni Dash and Am{\'{e}}lie Reymond and Emma S. Spiro and Aylin Caliskan},
title = {{Persona-Assigned Large Language Models Exhibit Human-Like Motivated Reasoning}},
booktitle = {{Findings of the Association for Computational Linguistics: ACL (ACL Findings)}},
pages = {12043-12069},
publisher = {ACL},
year = {2026}
}

@inproceedings{DMRKN23,
author = {Ameet Deshpande and Vishvak Murahari and Tanmay Rajpurohit and Ashwin Kalyan and Karthik Narasimhan},
title = {{Toxicity in chatgpt: Analyzing persona-assigned language models}},
booktitle = {{Findings of the Association for Computational Linguistics: EMNLP (EMNLP Findings)}},
pages = {1236-1270},
publisher = {ACL},
year = {2023}
}

@inproceedings{FL25,
author = {Neil Fasching and Yphtach Lelkes},
title = {{Model-Dependent Moderation: Inconsistencies in Hate Speech Detection Across LLM-based Systems}},
booktitle = {{Findings of the Association for Computational Linguistics: ACL (ACL Findings)}},
pages = {22271-22285},
publisher = {ACL},
year = {2025}
}

@inproceedings{FDCLBSVSK18,
author = {Antigoni{-}Maria Founta and Constantinos Djouvas and Despoina Chatzakou and Ilias Leontiadis and Jeremy Blackburn and Gianluca Stringhini and Athena Vakali and Michael Sirivianos and Nicolas Kourtellis},
title = {{Large Scale Crowdsourcing and Characterization of Twitter Abusive Behavior}},
booktitle = {{International Conference on Web and Social Media (ICWSM)}},
pages = {491-500},
publisher = {AAAI},
year = {2018}
}

@inproceedings{HGC23,
author = {Pengcheng He and Jianfeng Gao and Weizhu Chen},
title = {{DeBERTaV3: Improving DeBERTa using ELECTRA-Style Pre-Training with Gradient-Disentangled Embedding Sharing}},
booktitle = {{International Conference on Learning Representations (ICLR)}},
year = {2023}
}

@inproceedings{HKA23,
author = {Fan Huang and Haewoon Kwak and Jisun An},
title = {{Is ChatGPT better than Human Annotators? Potential and Limitations of ChatGPT in Explaining Implicit Hate Speech}},
booktitle = {{The Web Conference (WWW)}},
pages = {294-297},
publisher = {ACM},
year = {2023}
}

@inproceedings{KSG25,
author = {Gauri Kambhatla and Chantal Shaib and Venkata Govindarajan},
title = {{Measuring Lexical Diversity of Synthetic Data Generated through Fine-Grained Persona Prompting}},
booktitle = {{Findings of the Association for Computational Linguistics: EMNLP (EMNLP Findings)}},
pages = {21024-21033},
publisher = {ACL},
year = {2025}
}

@inproceedings{KSYVLWGLWN25,
author = {Seungone Kim and Juyoung Suk and Xiang Yue and Vijay Viswanathan and Seongyun Lee and Yizhong Wang and Kiril Gashteovski and Carolin Lawrence and Sean Welleck and Graham Neubig},
title = {{Evaluating Language Models as Synthetic Data Generators}},
booktitle = {{Annual Meeting of the Association for Computational Linguistics (ACL)}},
pages = {6385-6403},
publisher = {ACL},
year = {2025}
}

@inproceedings{KW13,
author = {Irene Kwok and Yuzhou Wang},
title = {{Locate the Hate: Detecting Tweets against Blacks}},
booktitle = {{AAAI Conference on Artificial Intelligence (AAAI)}},
pages = {1621-1622},
publisher = {AAAI},
year = {2013}
}

@inproceedings{LDF24,
author = {Andy Liu and Mona T. Diab and Daniel Fried},
title = {{Evaluating Large Language Model Biases in Persona-Steered Generation}},
booktitle = {{Findings of the Association for Computational Linguistics: ACL (ACL Findings)}},
pages = {9832-9850},
publisher = {ACL},
year = {2024}
}

@inproceedings{LDZH22,
author = {Florian Ludwig and Klara Dolos and Torsten Zesch and Eleanor Hobley},
title = {{Improving Generalization of Hate Speech Detection Systems to Novel Target Groups via Domain Adaptation}},
booktitle = {{Workshop on Online Abuse and Harms (WOAH)}},
pages = {29-39},
publisher = {ACL},
year = {2022}
}

@inproceedings{MZANLAJW23,
author = {Todor Markov and Chong Zhang and Sandhini Agarwal and Florentine Eloundou Nekoul and Theodore Lee and Steven Adler and Angela Jiang and Lilian Weng},
title = {{A Holistic Approach to Undesired Content Detection in the Real World}},
booktitle = {{AAAI Conference on Artificial Intelligence (AAAI)}},
pages = {15009-15018},
publisher = {AAAI},
year = {2023}
}

@inproceedings{NTTMC16,
author = {Chikashi Nobata and Joel R. Tetreault and Achint Thomas and Yashar Mehdad and Yi Chang},
title = {{Abusive Language Detection in Online User Content}},
booktitle = {{International Conference on World Wide Web (WWW)}},
pages = {145-153},
publisher = {ACM},
year = {2016}
}

@inproceedings{QBLBW19,
author = {Jing Qian and Anna Bethke and Yinyin Liu and Elizabeth M. Belding and William Yang Wang},
title = {{A Benchmark Dataset for Learning to Intervene in Online Hate Speech}},
booktitle = {{Conference on Empirical Methods in Natural Language Processing and International Joint Conference on Natural Language Processing (EMNLP-IJCNLP)}},
pages = {4754-4763},
publisher = {ACL},
year = {2019}
}

@inproceedings{SCC19,
author = {Koustuv Saha and Eshwar Chandrasekharan and Munmun De Choudhury},
title = {{Prevalence and Psychological Effects of Hateful Speech in Online College Communities}},
booktitle = {{ACM Conference on Web Science (WebSci)}},
pages = {255-264},
publisher = {ACM},
year = {2019}
}

@inproceedings{TLMHFOR25,
author = {Manuel Tonneau and Diyi Liu and Niyati Malhotra and Scott A. Hale and Samuel Fraiberger and V{\'{\i}}ctor Orozco{-}Olvera and Paul R{\"{o}}ttger},
title = {{HateDay: Insights from a Global Hate Speech Dataset Representative of a Day on Twitter}},
booktitle = {{Annual Meeting of the Association for Computational Linguistics (ACL)}},
pages = {2297-2321},
publisher = {ACL},
year = {2025}
}

@inproceedings{WH12,
author = {William Warner and Julia Hirschberg},
title = {{Detecting Hate Speech on the World Wide Web}},
booktitle = {{Workshop on Language in Social Media (LSM)}},
pages = {19-26},
publisher = {ACL},
year = {2012}
}

@inproceedings{WH16,
author = {Zeerak Waseem and Dirk Hovy},
title = {{Hateful Symbols or Hateful People? Predictive Features for Hate Speech Detection on Twitter}},
booktitle = {{NAACL Student Research Workshop (NSRW)}},
pages = {88-93},
publisher = {ACL},
year = {2016}
}

@inproceedings{BPV24,
author = {Nikolaj Bauer and Moritz Preisig and Martin Volk},
title = {{Offensiveness, Hate, Emotion and {GPT}: Benchmarking {GPT}3.5 and {GPT}4 as Classifiers on Twitter-specific Datasets}},
booktitle = {{Workshop on Threat, Aggression \& Cyberbullying (TRAC)}},
pages = {126-133},
publisher = {ELRA and ICCL},
year = {2024}
}

@article{JDJ21,
author = {Jeff Johnson and Matthijs Douze and Herv{\'e} J{\'e}gou},
title = {{Billion-scale Similarity Search with GPUs}},
journal = {{IEEE Transactions on Big Data}},
publisher = {IEEE},
year = {2021}
}

@article{APP22,
author = {Aym{\'{e}} Arango and Jorge P{\'{e}}rez and Barbara Poblete},
title = {{Hate speech detection is not as easy as you may think: {A} closer look at model validation (extended version)}},
journal = {{Information Systems}},
publisher = {Elsevier},
year = {2022}
}

@article{CG26,
author = {Paolo Climaco and Jochen Garcke},
title = {{Density-Aware Farthest Point Sampling}},
journal = {{Transactions on Machine Learning Research}},
year = {2026}
}

@article{FN18,
author = {Paula Fortuna and S{\'{e}}rgio Nunes},
title = {{A Survey on Automatic Detection of Hate Speech in Text}},
journal = {{ACM Computing Surveys}},
publisher = {ACM},
year = {2018}
}

@article{GHC25,
author = {Adri{\'{a}}n Gir{\'{o}}n and Javier Huertas{-}Tato and David Camacho},
title = {{{LLM} synthetic generation to enhance online content moderation generalization in hate speech scenarios}},
journal = {{Computing}},
publisher = {Springer},
year = {2025}
}

@article{G85,
author = {Teofilo F. Gonzalez},
title = {{Clustering to Minimize the Maximum Intercluster Distance}},
journal = {{Theoretical Computer Science}},
publisher = {Elsevier},
year = {1985}
}

@article{PBSBP21,
author = {Fabio Poletto and Valerio Basile and Manuela Sanguinetti and Cristina Bosco and Viviana Patti},
title = {{Resources and benchmark corpora for hate speech detection: a systematic review}},
journal = {{Language Resources and Evaluation}},
publisher = {Springer},
year = {2021}
}

@article{SWYCCH24,
author = {Zhouxing Shi and Yihan Wang and Fan Yin and Xiangning Chen and Kai{-}Wei Chang and Cho{-}Jui Hsieh},
title = {{Red Teaming Language Model Detectors with Language Models}},
journal = {{Transactions of the Association for Computational Linguistics}},
publisher = {The MIT Press},
year = {2024}
}

@article{SLSVLM25,
author = {Dvora Shmulewitz and Maor Daniel Levitin and Vera Skvirsky and Merav Vider and Shaul Lev{-}Ran and Mario Mikulincer},
title = {{Exposure to online hate speech is positively associated with post-traumatic stress disorder symptom severity}},
journal = {{Scientific Reports}},
publisher = {Nature Publishing Group},
year = {2025}
}

@article{LOGDJCLLZS19,
author = {Yinhan Liu and Myle Ott and Naman Goyal and Jingfei Du and Mandar Joshi and Danqi Chen and Omer Levy and Mike Lewis and Luke Zettlemoyer and Veselin Stoyanov},
title = {{RoBERTa: {A} Robustly Optimized {BERT} Pretraining Approach}},
journal = {{CoRR abs/1907.11692}},
year = {2019}
}

@article{KBSV20,
author = {Chris J. Kennedy and Geoff Bacon and Alexander Sahn and Claudia von Vacano},
title = {{Constructing interval variables via faceted Rasch measurement and multitask deep learning: a hate speech application}},
journal = {{CoRR abs/2009.10277}},
year = {2020}
}

@article{ZWCNKF23,
author = {Andy Zou and Zifan Wang and Nicholas Carlini and Milad Nasr and J. Zico Kolter and Matt Fredrikson},
title = {{Universal and Transferable Adversarial Attacks on Aligned Language Models}},
journal = {{CoRR abs/2307.15043}},
year = {2023}
}

@article{T243,
author = {Qwen Team},
title = {{Qwen2.5 Technical Report}},
journal = {{CoRR abs/2412.15115}},
year = {2024}
}

@article{GCWYMY24,
author = {Tao Ge and Xin Chan and Xiaoyang Wang and Dian Yu and Haitao Mi and Dong Yu},
title = {{Scaling Synthetic Data Creation with 1,000,000,000 Personas}},
journal = {{CoRR abs/2406.20094}},
year = {2024}
}

@article{GDF25,
author = {Faeze Ghorbanpour and Daryna Dementieva and Alexander Fraser},
title = {{Can Prompting LLMs Unlock Hate Speech Detection across Languages? {A} Zero-shot and Few-shot Study}},
journal = {{CoRR abs/2505.06149}},
year = {2025}
}

@article{MFC20,
author = {Marzieh Mozafari and Reza Farahbakhsh and No{\"{e}}l Crespi},
title = {{Hate Speech Detection and Racial Bias Mitigation in Social Media based on {BERT} model}},
journal = {{CoRR abs/2008.06460}},
year = {2020}
}

@article{VD20,
author = {Bertie Vidgen and Leon Derczynski},
title = {{Directions in Abusive Language Training Data: Garbage In, Garbage Out}},
journal = {{CoRR abs/2004.01670}},
year = {2020}
}

@article{WHADKM25,
author = {Manuel Weber and Moritz Huber and Maximilian Auch and Alexander D{\"{o}}schl and Max{-}Emanuel Keller and Peter Mandl},
title = {{Digital Guardians: Can GPT-4, Perspective API, and Moderation {API} reliably detect hate speech in reader comments of German online newspapers?}},
journal = {{CoRR abs/2501.01256}},
year = {2025}
}

@article{YZ21,
author = {Wenjie Yin and Arkaitz Zubiaga},
title = {{Towards generalisable hate speech detection: a review on obstacles and solutions}},
journal = {{CoRR abs/2102.08886}},
year = {2021}
}

@book{C77,
author = {William G. Cochran},
title = {{Sampling Techniques}},
publisher = {John Wiley \& Sons},
year = {1977}
}

@misc{4chan,
title = {{4chan}},
author = {{4chan community support LLC}},
howpublished = {\url{https://www.4chan.org/}},
}

@misc{Perspective,
title = {{Perspective API}},
author = {{Perspective}},
year = {2023},
howpublished = {\url{https://www.perspectiveapi.com}},
}

@misc{Detoxify,
title = {{Detoxify}},
author = {Laura Hanu and {Unitary team}},
howpublished = {\url{https://github.com/unitaryai/detoxify}},
}

@misc{Vicuna,
title = {{Vicuna: An Open-Source Chatbot Impressing GPT-4 with 90\%* ChatGPT Quality}},
author = {The Vicuna Team},
year = {2023},
howpublished = {\url{https://lmsys.org/blog/2023-03-30-vicuna/}},
}

@misc{claude,
title = {{Claude}},
author = {Anthropic},
howpublished = {\url{https://claude.ai/}},
}

@misc{llama3_1_blog,
title = {{Introducing Llama 3.1: Our most capable models to date}},
author = {{Meta}},
year = {2024},
howpublished = {\url{https://ai.meta.com/blog/meta-llama-3-1/}},
}

@misc{llama3_2_blog,
title = {{Llama 3.2: Revolutionizing edge AI and vision with open, customizable models}},
author = {{Meta}},
year = {2024},
howpublished = {\url{https://ai.meta.com/blog/llama-3-2-connect-2024-vision-edge-mobile-devices/}},
}

@misc{GPT4o,
title = {{Hello GPT-4o}},
author = {OpenAI},
year = {2024},
howpublished = {\url{https://openai.com/index/hello-gpt-4o/}},
}

@misc{OpenAI_Moderation,
title = {{OpenAI Moderation API}},
author = {OpenAI},
howpublished = {\url{https://developers.openai.com/api/docs/guides/moderation}},
}

@misc{gemini_pro,
title = {{Gemini 2.5 Pro}},
author = {{Google}},
year = {2025},
howpublished = {\url{https://docs.cloud.google.com/vertex-ai/generative-ai/docs/models/gemini/2-5-pro}},
}

@misc{gemini_flash,
title = {{Gemini Flash}},
author = {{Google}},
howpublished = {\url{https://docs.cloud.google.com/vertex-ai/generative-ai/docs/models/gemini/2-5-flash}},
}

@misc{GPT-4o-mini,
title = {{GPT‑4o mini: advancing cost-efficient intelligence}},
author = {OpenAI},
year = {2024},
howpublished = {\url{https://openai.com/index/gpt-4o-mini-advancing-cost-efficient-intelligence/}},
}

@misc{claude4,
title = {{Introducing Claude 4}},
author = {Anthropic},
year = {2025},
howpublished = {\url{https://www.anthropic.com/news/claude-4}},
}

@misc{cardiffnlp_hate_latest,
title = {{cardiffnlp/twitter-roberta-base-hate-latest}},
author = {{Cardiff NLP}},
year = {2023},
howpublished = {\url{https://huggingface.co/cardiffnlp/twitter-roberta-base-hate-latest}},
}

@misc{llama_guard_3,
title = {{Llama Guard 3}},
author = {Meta},
year = {2024},
howpublished = {\url{https://www.llama.com/docs/model-cards-and-prompt-formats/llama-guard-3/}},
}

@misc{UN_hate_speech,
title = {{What is hate speech?}},
author = {{United Nations}},
howpublished = {\url{https://www.un.org/en/hate-speech/understanding-hate-speech/what-is-hate-speech}},
}

@misc{unesco_ipsos_2023_news,
title = {{Social Media Feeds widest source of disinformation \& fake news - UNESCO-Ipsos Survey}},
author = {Madhurima Bhatia},
year = {2023},
howpublished = {\url{https://www.ipsos.com/en-in/social-media-feeds-widest-source-disinformation-fake-news-unesco-ipsos-survey}},
}

@misc{claude_opus_46,
title = {{Introducing Claude Opus 4.6}},
author = {{Anthropic}},
year = {2026},
howpublished = {\url{https://www.anthropic.com/news/claude-opus-4-6}},
}

@misc{deepseek_v4_flash,
title = {{DeepSeek V4 Preview Release}},
author = {{DeepSeek}},
year = {2026},
howpublished = {\url{https://api-docs.deepseek.com/news/news260424/}},
}

@misc{qwen3,
title = {{Qwen3: Think Deeper, Act Faster}},
author = {{Qwen Team}},
year = {2025},
howpublished = {\url{https://qwenlm.github.io/blog/qwen3/}},
}

@misc{mistral_small_3,
title = {{Mistral Small 3}},
author = {{Mistral AI}},
year = {2025},
howpublished = {\url{https://mistral.ai/news/mistral-small-3/}},
}

@misc{memes,
title = {{Internet Memes}},
author = {Wikipedia},
howpublished = {\url{https://en.wikipedia.org/wiki/Internet_meme}},
}

@misc{gemini-2.5-flash,
author = {Google},
year = {2025},
howpublished = {\url{https://ai.google.dev/gemini-api/docs/models/gemini-2.5-flash}},
}

\appendix

\section{Open Science}

We provide the code and supporting artifacts on GitHub at
\url{https://github.com/TrustAIRLab/PersonaHate}.
The associated dataset repository is hosted on Hugging Face at
\url{https://huggingface.co/datasets/TrustAIRLab/PersonaHate}.

The release includes:
\begin{itemize}
    \item Code for the \method data synthesis pipeline, including persona selection, prompt construction, generation orchestration, refusal filtering, and multi-judge annotation.
    \item Code for detector fine-tuning and benchmark evaluation.
    \item The selected persona set used in our experiments to support reproduction of the key pipeline stages.
    \item A representative subset of the group-balanced training data.
\end{itemize}

The full generated corpus contains 791K valid samples, and the curated group-balanced training set contains 67K samples.

These artifacts are intended for research use, including reproducing our experiments and developing hate speech detection systems.
As the released data includes LLM-generated hate speech targeting specific identity groups, reader discretion is recommended.

\section{Ethical Considerations}
\label{app:ethics}

This work studies persona-driven synthetic data generation for hate speech detection.
Because it involves generating and analyzing hateful or offensive language, we consider potential risks to affected identity groups, researchers, downstream users, and the broader public.
Our goal is not to release a general-purpose hate-speech generator, but to construct large-scale and diverse training data for improving hate speech detectors.

Our pipeline does not collect private user data.
The generated samples are synthetic and are produced by LLMs conditioned on persona descriptions, identity groups, and sentiment directions.
The large-scale annotation process is performed by the six-judge pipeline and moderation tools, with only limited manual inspection by the authors for quality control.
For the public benchmark datasets used in evaluation, we follow their intended research use and report results in aggregate.

The main ethical concern is the dual-use risk of scalable hate-speech generation.
The underlying LLMs used in our pipeline are already publicly accessible; thus, the core generation capability is not introduced by \method.
Our contribution is to systematize this capability for safety research through candidate data generation, filtering, six-judge annotation, and downstream detector training.
At the same time, we acknowledge that packaging these components into a scalable synthesis pipeline may introduce additional misuse risk.

The empirical results support this intended use.
\method does not merely improve detection of synthetic hate speech; it also improves detection on human-authored benchmarks, especially for smaller deployable models.
This indicates that the generated data can strengthen practical hate-speech detection rather than only improve synthetic-pattern recognition.

There is also a risk that identity-group labels or persona-conditioned generation could reinforce, or be interpreted as reinforcing, harmful stereotypes.
We therefore use identity-group labels only as detection targets and evaluation attributes.
They are not used to profile individuals, make claims about protected groups, or justify stereotypes.
We also avoid including unnecessary hateful examples in the paper and provide a content warning for readers.

To reduce misuse risk, the dataset and code will be released under responsible-use terms.
We will not provide an unrestricted hate-generation interface.
Prompt templates will be sanitized where necessary, and the documentation will clearly specify hate-speech detector training and safety evaluation as the intended uses.

\section{Human Evaluation Details}
\label{app:human_eval}

\subsection{Sample Size Determination}

We determine the sample size following standard confidence-interval analysis for proportions~\cite{C77}.
For a binary proportion, the required sample size is estimated as
\[
n=\frac{z_{1-\alpha/2}^{2}p(1-p)}{\epsilon^{2}},
\]
where $z_{1-\alpha/2}$ denotes the critical value for the desired confidence level, $p$ is the expected proportion, and $\epsilon$ is the target margin of error.
With a 95\% confidence level ($\alpha=0.05$, $z_{1-\alpha/2}=1.96$), a target margin of error of $\epsilon=0.05$, and the conservative assumption $p=0.5$, which maximizes the required sample size, we obtain
\[
n=\frac{1.96^{2}\times0.5\times0.5}{0.05^{2}}\approx384.2.
\]
Accordingly, we use 400 samples for each human evaluation.
Samples are presented in randomized order, and annotators independently complete the assigned annotation tasks without access to other annotators' judgments.

\subsection{Multi-Judge Label Reliability}

We evaluate whether the labels produced by our six-judge annotation pipeline align with human judgments.
The \method samples are stratified across different levels of six-judge vote agreement so that the evaluation covers both high-consensus and ambiguous cases.
For each sample, annotators independently classify the text as hate or non-hate following the same binary hate-speech definition used in our annotation pipeline.
We aggregate the human annotations by majority vote and compare the resulting human-majority label with the label assigned by our $\geq4/6$ voting rule.

The human-majority labels agree with the \method labels at 87.0\% accuracy, 89.3\% F1, and Cohen's $\kappa=0.727$.
The average pairwise agreement among human annotators is $\kappa=0.411$.
Among the 281 samples for which all three annotators assign the same label, \method agrees with the unanimous human judgment on 276 samples (98.2\%).
Agreement is also higher for cases with strong six-judge consensus; for example, human agreement reaches 93\% for samples receiving 6/6 hate votes and 96\% for samples receiving 0/6 hate votes.
These results indicate that the aggregated six-judge labels align closely with human consensus, particularly for clear and high-confidence cases.

\subsection{Persona Fidelity and Stereotypicality}

We evaluate whether personas extracted from 4chan /pol/ posts faithfully reflect their source texts and whether the extraction process introduces stereotypical characterization.
For each source-post--persona pair, annotators rate how faithfully the persona reflects the attitudes, communication style, and characteristics expressed in the source post on a 1--5 Likert scale, where 5 indicates the highest fidelity.
Annotators additionally provide a binary judgment indicating whether the extracted persona contains stereotypical characterization.

The extracted personas receive an overall mean fidelity score of 4.20/5, and 85.4\% of the ratings are at least 4/5.
Only 3.2\% of the evaluated personas are flagged as stereotypical.
These results indicate that the extraction procedure generally preserves the characteristics expressed in the source posts, with limited evidence of stereotypical characterization.

\subsection{Generated-Text Naturalness}

We evaluate whether \method-generated texts resemble naturally occurring human-written content by comparing them with human-written samples from HateXplain.
The source identity of each text is hidden from annotators, and the samples are presented in randomized order.
For each text, annotators predict whether it is AI-generated or human-written and then rate its naturalness on a 1--5 Likert scale, where 5 indicates the highest naturalness.

The \method samples receive an average naturalness score of 4.06/5, compared with 4.47/5 for human-written HateXplain samples.
The difference in naturalness is statistically significant ($p<0.001$), although both sets receive high absolute naturalness scores.
Moreover, 59.0\% of \method samples are classified as human-written, while 77.5\% of human-written HateXplain samples are correctly recognized as human-written.
The overall source-classification accuracy is 59.2\%.
These results indicate that persona-conditioned generation produces highly natural synthetic text and that a substantial fraction of the generated samples are difficult for human annotators to distinguish from human-written content.

\section{Persona Extraction Examples}
\label{app:persona_examples}

Table~\ref{tab:persona_examples} shows six examples of hateful personas extracted from 4chan /pol/ posts. 
The posts span a range of identity attack scores (0.71 to 0.97) and cover different persona archetypes: from overtly aggressive individuals using extreme profanity (top rows) to politically frustrated users expressing subtler forms of prejudice (bottom rows). 
For each post, GPT-4o infers a persona description that captures the author's attitudes, communication style, and ideological leanings. 
These persona descriptions serve as character profiles for downstream hate speech generation.

\noindent\fbox{%
  \parbox{0.95\linewidth}{%
    \textbf{System prompt:} 

Analyze who would write this text. \\
IMPORTANT: Start your response with the persona description directly. \\
DO NOT start with phrases like: \\
- "The text is likely to be written by" \\
- "This text appears to be written by" \\
- "The author is likely" \\
Just describe the persona directly. \\

    \textbf{User prompt:} 

Based on this text, describe the persona who would write it: \\
\{text\} \\
Remember: Start your description directly without any introductory phrases!
  }%
}

\begin{table*}[t!]
\centering
\caption{Examples of hateful personas extracted from 4chan /pol/ posts via GPT-4o. Each row shows the post ID, its Perspective API identity attack score, the source post, and the inferred persona description. Content warning: source posts contain offensive language.}
\label{tab:persona_examples}
\small
\setlength{\tabcolsep}{5pt}
\begin{tabular}{clp{0.42\textwidth}p{0.42\textwidth}}
\toprule
\textbf{Tox.} & \textbf{Post ID} & \textbf{Source Post} & \textbf{Extracted Persona} \\
\midrule
0.97 & 110938075 & Shut the f**k up, dude. I don't care that you're right I just don't want any f**king American to ever respond to me in ANY f**king thread. Do you hear me you ``son of bubblegum b***h'' or whatever you f**king say in that s**t hole. By now you're probably thinking: ``hurr *eats cheetohs* ders no way dis is real it has 2 b *sips 3 litre mnt dew* copy pasta!''. Well it isn't and you're wrong. This is all me, baby. Calling you out for what you are you sack of ``owl s**t'' or whatever the f**k. Don't even think to reply to this, or infact, ANY f**king Australian poster in ANY thread because I guarantee you're going to feel the full brunt of my steel-capped boot slamming down on your fat f**king skull. Enjoy your day, c**t. & An aggressive and confrontational individual with a strong disdain for Americans, likely stemming from cultural or personal grievances. This persona is unapologetically rude and uses profane language to express anger and superiority. They possess a volatile temperament and possibly harbor national pride, as evidenced by the explicit disapproval of American responses to Australian posts. The writer adopts an internet-savvy tone, referencing memes and stereotypes about Americans, indicating familiarity with online culture and possibly frequent participation in online forums or discussions. This person is confrontational and enjoys using intimidation tactics to assert dominance over perceived adversaries. \\
\addlinespace
0.84 & 92427541 & I'm with Alex on this one. If Hillary wins, f**k this country. It was a good run, but the US really is prepping itself for the garbage bin of history. Anyone know some countries (that aren't cucked) which allow citizenship through military service? Or maybe that have low skilled work visas until citizenship attainment? I'll pick cotton, I really don't give a f**k. I have couple degrees, but I doubt New Zealand, Australia, Medpackland, or wherever else gives a s**t about a Law Enforcement \& US law degree, so yeah I think I'm basically f**ked when it comes to being regarded as a ``skilled'' immigrant. & A disgruntled and politically frustrated individual with a strong opposition to Hillary Clinton and a pessimistic view of the future of the United States. They feel disillusioned with the political direction of the country and are exploring options for emigration. They have a background in law enforcement and US law but feel that their qualifications may not be valued internationally. Despite their educational background, they express a willingness to take on low-skilled work to obtain citizenship in another country. \\
\addlinespace
0.73 & 96282744 & Yeah, they tell you that, but you need to look at their actions. Lefties are constantly complaining about the 1\%, yet they suck the d**k of every actor, musician, artist, etc., and follow trends more than any other group of people. As long as it's some millionaire with a background in ``art'' rather than banking telling them what to do, they won't bat an eye. They'll sit there and say ``Well Hillary is a politician, but all of these good-hearted people like Leonardo DiCaprio are voting for her, so she must be okay''. & A politically conservative individual with a critical view of leftist ideologies and behaviors. They express skepticism towards the perceived hypocrisy of left-leaning individuals, particularly regarding issues of wealth and celebrity influence. This person may feel frustrated with what they see as inconsistencies in the values and actions of those on the left, especially when it comes to idolizing wealthy individuals in the entertainment industry while criticizing wealth accumulation in other sectors. The tone suggests a tendency towards sarcasm and a desire to challenge the credibility and decision-making of liberal individuals or groups. \\
\addlinespace
0.71 & 94529252 & Watching the last season was the worst -- every episode is Rick (the supposed alpha) kowtowing to the needs of every woman he possibly can. That last mission was to get maggie to the `doctor' for her cramp, yet she was fine at the end of this episode. They abandoned the safety of their walls/guns/ammo to get captured because a b***h has a p***y ache. Rewatch some of last season, literally every episode is him getting overruled by carol and maggie and then his idea getting overruled and him just forgetting that he had a plan. Rick literally destroyed a family because the husband slapped his wife. And what do these people do all day? Why is there crops? No potato patches, no corn, no pig stye, no nothing, it's just the females sitting around doing nothing in their electricity powered houses waiting until the real men show up and kill them all. & A disgruntled viewer of a television series, possibly ``The Walking Dead,'' who feels frustrated with the show's depiction of gender dynamics and character decisions. This person appears to have strong opinions about traditional gender roles, expressing disdain for what they perceive as male characters being subservient to female characters. They seem to be critical of narrative choices that prioritize female characters' needs or perspectives and are frustrated by perceived plot inconsistencies, such as characters' survival tactics and resource management. The tone suggests they are passionate, possibly using hyperbolic language to emphasize their displeasure with the show's direction. \\
\bottomrule
\end{tabular}
\end{table*}

\section{Persona Deduplication}
\label{app:deduplication}

The raw 4chan-derived personas contain substantial redundancy, since many posts express similar beliefs or rhetorical patterns.
We therefore conduct deduplication: given persona embeddings, we greedily retain a persona only if its cosine similarity to all previously retained personas is below a threshold $\tau=0.9$.
Specifically, given a persona pool $\mathcal{P} = \{p_1, \ldots, p_N\}$, we compute normalized embeddings $\mathbf{e}_i = \phi(p_i) / \|\phi(p_i)\|$ using \texttt{text-embedding-3-small}. 
We then apply greedy deduplication by initializing a retained set $\mathcal{S} = \emptyset$ and scanning each persona in sequence. 
Persona $p_i$ is retained if and only if
\begin{equation}
    \max_{j \in \mathcal{S}} \mathbf{e}_i^\top \mathbf{e}_j < \tau.
\end{equation}
This ensures that no two retained personas exceed the cosine similarity threshold. 
To accelerate nearest neighbor search, we maintain a Faiss inner product index~\cite{JDJ21} that is incrementally updated as new personas are retained. 
After deduplication, the hateful persona pool is reduced from 305,616 to 257,473 personas, removing 15.7\% of the original entries.

\section{Persona Selection Visualization}
\label{app:fps_umap}

\begin{figure*}[t!]
\centering
\includegraphics[width=\textwidth]{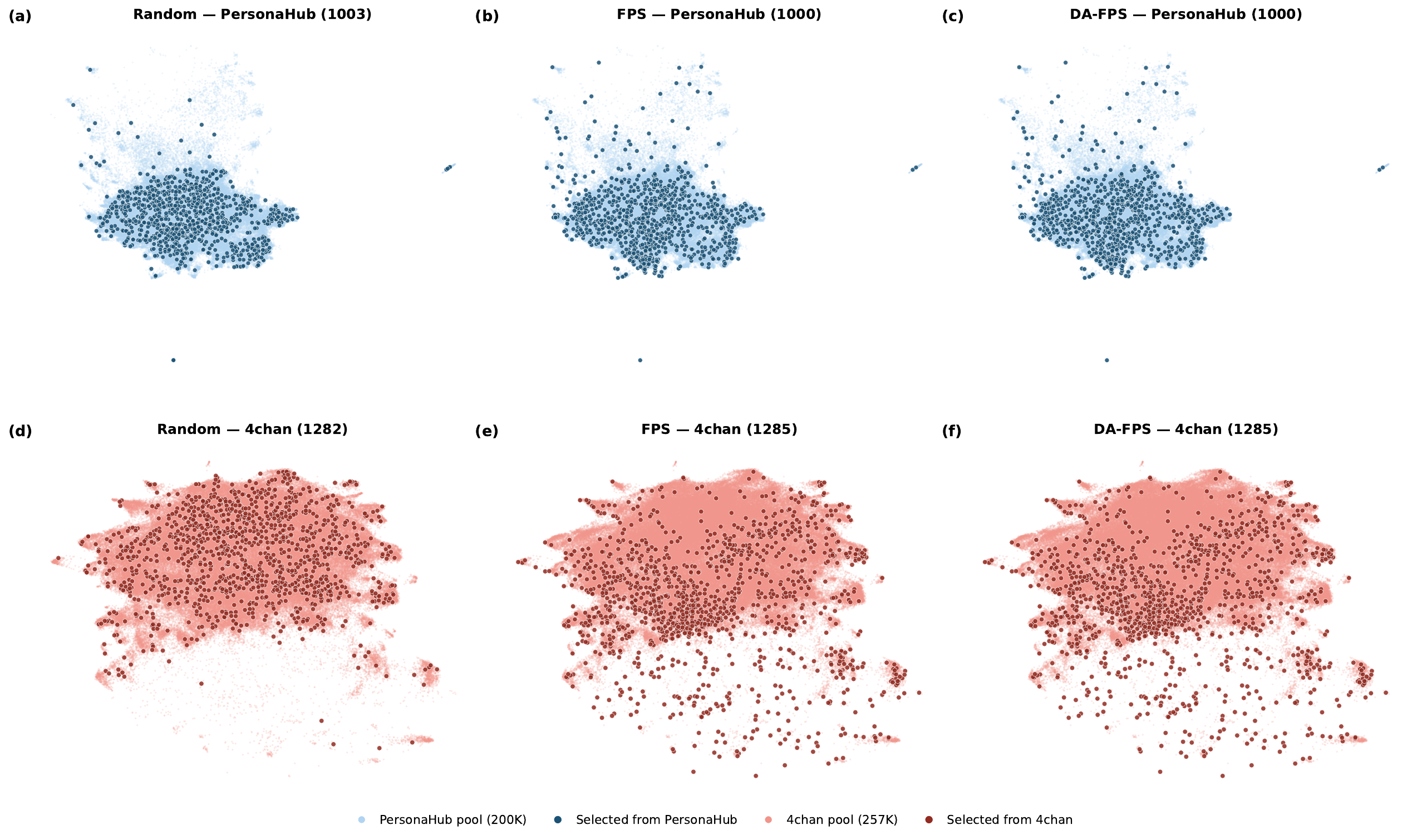}
\caption{UMAP visualization of persona selection methods. \textbf{Top row}: PersonaHub pool (200K personas). \textbf{Bottom row}: 4chan pool (257K personas). Columns show (a,\,d)~random sampling, (b,\,e)~FPS, and (c,\,f)~DA-FPS. Light points are the full pool; dark points are the selected subset. FPS and DA-FPS achieve uniform spatial coverage, while random sampling over-represents dense clusters.}
\label{fig:fps_umap}
\end{figure*}

Figure~\ref{fig:fps_umap} visualizes the persona selections produced by random sampling, FPS, and DA-FPS using UMAP projections of the full persona embedding space (457K points). 
Each row corresponds to one persona source; each column to a selection method. 
The background shows the complete pool for that source, and dark points indicate the selected personas.

Random sampling (a, d) concentrates selections in high-density regions and leaves sparse areas uncovered. 
Both FPS (b, e) and DA-FPS (c, f) spread selections uniformly across the embedding space, covering boundary regions that random sampling misses. 
The two methods produce visually indistinguishable results, consistent with their identical fill distance (Table~\ref{tab:fps_comparison}).

\begin{table}[t!]
\centering
\caption{Within-model, within-group semantic diversity (avg pairwise cosine distance of Sentence-BERT embeddings, averaged across 34 groups).
\method models show stable diversity; HateBenchSet models vary widely due to the absence of persona conditioning.}
\label{tab:within_model_diversity}
\small
\setlength{\tabcolsep}{5pt}
\begin{tabular}{lc|lc}
\toprule
\multicolumn{2}{c|}{\textbf{\method}} & \multicolumn{2}{c}{\textbf{HateBenchSet}} \\
\textbf{Model} & \textbf{Div.} & \textbf{Model} & \textbf{Div.} \\
\midrule
Gemini-2.5-Flash & 0.577 & OPT       & 0.758 \\
DeepSeek-R1      & 0.576 & Dolly     & 0.565 \\
Gemma-2-9B       & 0.544 & Baichuan  & 0.481 \\
Claude-3-Haiku   & 0.541 & Vicuna    & 0.469 \\
Qwen-2.5-7B      & 0.521 & GPT-3     & 0.399 \\
Llama-3.1-8B     & 0.521 & GPT-4     & 0.373 \\
GPT-4o-mini      & 0.473 &           & \\
Mistral-7B       & 0.449 &           & \\
\midrule
\textbf{Avg $\pm$ std} & \textbf{0.525 $\pm$ 0.046} & \textbf{Avg $\pm$ std} & \textbf{0.507 $\pm$ 0.131} \\
\bottomrule
\end{tabular}
\end{table}

\begin{table}[!htbp]
\centering
\caption{Structural and lexical diversity comparison between \method and HateBenchSet. Lexical metrics are computed on equal-sized samples (7{,}838 texts each).}
\label{tab:diversity_vs_hatebench}
\small
\setlength{\tabcolsep}{5pt}
\begin{tabular}{lcc}
\toprule
\textbf{Metric} & \textbf{\method} & \textbf{HateBenchSet} \\
\midrule
Total samples           & 791{,}283           & 7{,}838       \\
Generators              & 8                   & 6             \\
Unique personas/prompts & 2{,}285 / 77{,}690  & --- / 204     \\
Identity groups         & 34                  & 34            \\
\midrule
Distinct-1              & 0.064               & 0.051         \\
Distinct-2              & 0.345               & 0.325         \\
Distinct-3              & 0.656               & 0.634         \\
Vocabulary size         & 32{,}976            & 27{,}132      \\
Avg length (tokens)     & $65 \pm 33$         & $68 \pm 55$   \\
\bottomrule
\end{tabular}
\end{table}

\begin{table}[t!]
\centering
\caption{The 34 identity groups used for speech generation, organized into six categories.}
\label{tab:identity_groups}
\small
\setlength{\tabcolsep}{5pt}
\begin{tabular}{ll}
\toprule
\textbf{Category} & \textbf{Identity Groups} \\
\midrule
Race/Ethnicity & Asian, Black or African American, \\
(7) & Latino or Non-White Hispanic, Middle Eastern, \\
 & Native American or Alaska Native, \\
 & Pacific Islander, Non-Hispanic White \\
\midrule
Religion & Atheists, Buddhists, Christians, Hindus, \\
(7) & Jews, Mormons, Muslims \\
\midrule
Immigration/ & Immigrants, Migrant Workers, \\
Origin (5) & People From a Specific Country, \\
 & Undocumented People, Refugees \\
\midrule
Gender & Men, Women, Non-Binary or Third Gender, \\
(6) & Transgender Men, Transgender Women, \\
 & Transgender (Unspecified) \\
\midrule
Sexual & Bisexual, Gay, Lesbian, Heterosexual \\
Orientation (4) & \\
\midrule
Disability & Physical Disabilities, Cognitive Disorders \\
(5) & or Learning Disability, Mental Health \\
 & Problems, Visually Impaired, \\
 & Hearing Impaired \\
\bottomrule
\end{tabular}
\end{table}

\begin{table}[t!]
\centering
\caption{Refusal rates under direct vs.\ persona-based prompting across three API-gated models. Each condition uses 100 personas $\times$ 34 groups = 3{,}400 prompts.}
\label{tab:refusal_cross_model}
\small
\setlength{\tabcolsep}{5pt}
\begin{tabular}{lccc}
\toprule
\textbf{Model} & \textbf{Direct (\%)} & \textbf{Persona (\%)} & \textbf{$\Delta$ (pp)} \\
\midrule
GPT-4o-mini       & 96.2 & 29.5 & $-$66.7 \\
Gemini-2.5-Flash  & 100.0 & 79.9 & $-$20.1 \\
Claude-3-Haiku    & 100.0 & 87.7 & $-$12.3 \\
\bottomrule
\end{tabular}
\end{table}

\section{Identity Groups}
\label{app:identity_groups}

Table~\ref{tab:identity_groups} shows the 34 identity groups covering six categories.
Compared with MHS, we omit the age categories following HateBenchSet, as hate speech targeting an age group is hard to define.

\section{\method Pipeline Statistics}
\label{app:dataset_overview}

Table~\ref{tab:dataset_overview} shows the overview of our \method pipeline and the resulting dataset.

\begin{table}[t!]
\centering
\caption{\method pipeline statistics.
Compliance rate is the fraction of prompts that yield valid non-refused outputs after post-processing.}
\label{tab:dataset_overview}
\small
\setlength{\tabcolsep}{5pt}
\begin{tabular}{lr}
\toprule
\textbf{Property} & \textbf{Value} \\
\midrule
Total prompts & 1{,}243{,}040 \\
Valid samples & 791{,}283 \\
Overall compliance rate & 63.7\% \\
\midrule
Hate ($\geq 4/6$) & 109{,}788 (13.9\%) \\
Non-hate ($\leq 3/6$) & 681{,}495 (86.1\%) \\
\midrule
Identity groups & 34 \\
Group categories & 6 \\
Unique personas & 2{,}285 \\
Generator models & 8 \\
Judge models & 6 \\
Voting threshold & $\geq 4/6$ \\
\midrule
Training set & 67{,}452 \\
\quad Hate / Non-hate & 33{,}726 / 33{,}726 \\
\bottomrule
\end{tabular}
\end{table}

\subsection{Generation Efficiency}
\label{app:generation_efficiency}

Table~\ref{tab:generation_efficiency} reports the detailed generation efficiency of each model, including wall-clock time, throughput, and total generated tokens.

\begin{table}[t!]
\centering
\caption{Generation efficiency per model.
Wall-clock time includes retries.
Throughput is measured in processed prompts per minute (p/m).
Output Tokens are the total input token amount (M).
}
\label{tab:generation_efficiency}
\small
\setlength{\tabcolsep}{5pt}
\begin{tabular}{lccc}
\toprule
\textbf{Model} & \textbf{Wall Time} & \textbf{Throughput} & \textbf{Output Tok.\ } \\
\midrule
GPT-4o-mini     & 5.3\,h  & 489  & 10.4 \\
Claude          & 13.4\,h & 193  & 2.1 \\
Gemini          & 22.9\,h & 113  & 1.8 \\
DeepSeek-R1-8B  & 9.1\,h  & 284  & 28.4 \\
Qwen-2.5-7B     & 7.5\,h  & 345  & 25.6 \\
Ministral-8B    & 3.2\,h  & 809  & 11.2 \\
Gemma-2-9B      & 3.3\,h  & 785  & 8.6 \\
Llama-3-8B      & 9.4\,h  & 275  & 39.3 \\
\bottomrule
\end{tabular}
\end{table}

\section{Per-Group Annotation Statistics}
\label{app:group_stats}

Table~\ref{tab:group_stats} reports the annotation statistics for each of the 34 identity groups in \method, sorted by hate rate (fraction labeled hate by $\geq\!4/6$ vote). The total sample count per group varies only due to differential refusal rates across models (ratio 1.3:1), confirming near-uniform coverage by construction. Hate rates range from 3.5\% (Atheists) to 27.0\% (Jews), with traditionally marginalized groups receiving higher hate rates. The ``Ambiguous'' column reports samples with 2--3 hate votes out of 6, which are retained as non-hate in the primary dataset.

\begin{table*}[t!]
\centering
\caption{Per-group annotation statistics on \method (791{,}283 samples, 6 judges, $\geq\!4/6$ hate threshold). Groups are sorted by hate rate. Hate ($\geq\!4/6$) and non-hate ($\leq\!1/6$) represent high-confidence labels; ambiguous samples (2--3/6) are retained as non-hate.}
\label{tab:group_stats}
\small
\setlength{\tabcolsep}{5pt}
\begin{tabular}{llrrrrr}
\toprule
\textbf{Category} & \textbf{Identity Group} & \textbf{Total} & \textbf{Hate ($\geq$4/6)} & \textbf{Non-hate ($\leq$1/6)} & \textbf{Ambiguous} & \textbf{Hate \%} \\
\midrule
\multicolumn{7}{l}{\textit{Religion}} \\
\quad & Jews                              & 20{,}461 & 5{,}515 & 13{,}128 & 1{,}818 & 27.0 \\
\quad & Muslims                           & 21{,}849 & 4{,}981 & 14{,}670 & 2{,}198 & 22.8 \\
\quad & Hindus                            & 22{,}327 & 4{,}521 & 15{,}969 & 1{,}837 & 20.2 \\
\quad & Mormons                           & 23{,}376 & 2{,}047 & 18{,}537 & 2{,}792 &  8.8 \\
\quad & Buddhists                         & 25{,}210 & 2{,}108 & 20{,}722 & 2{,}380 &  8.4 \\
\quad & Christians                        & 24{,}289 & 1{,}236 & 20{,}691 & 2{,}362 &  5.1 \\
\quad & Atheists                          & 25{,}141 &    872 & 22{,}224 & 2{,}045 &  3.5 \\
\midrule
\multicolumn{7}{l}{\textit{Race \& Ethnicity}} \\
\quad & Asian                             & 21{,}377 & 4{,}143 & 15{,}033 & 2{,}201 & 19.4 \\
\quad & Pacific Islander                  & 22{,}505 & 4{,}332 & 16{,}102 & 2{,}071 & 19.2 \\
\quad & Middle Eastern                    & 21{,}907 & 4{,}102 & 15{,}822 & 1{,}983 & 18.7 \\
\quad & Latino or Non-White Hispanic      & 21{,}501 & 3{,}869 & 15{,}277 & 2{,}355 & 18.0 \\
\quad & Native American or Alaska Native  & 22{,}598 & 3{,}940 & 16{,}313 & 2{,}345 & 17.4 \\
\quad & Black or African American         & 20{,}746 & 3{,}282 & 15{,}478 & 1{,}986 & 15.8 \\
\quad & Non-Hispanic White                & 22{,}430 & 2{,}943 & 15{,}644 & 3{,}843 & 13.1 \\
\midrule
\multicolumn{7}{l}{\textit{Gender}} \\
\quad & Women                             & 24{,}870 & 4{,}931 & 17{,}108 & 2{,}831 & 19.8 \\
\quad & Transgender Women                 & 22{,}659 & 3{,}915 & 16{,}127 & 2{,}617 & 17.3 \\
\quad & Transgender Men                   & 23{,}053 & 3{,}663 & 16{,}488 & 2{,}902 & 15.9 \\
\quad & Transgender (Unspecified)         & 23{,}170 & 1{,}744 & 18{,}860 & 2{,}566 &  7.5 \\
\quad & Non-Binary or Third Gender        & 25{,}318 & 1{,}338 & 20{,}209 & 3{,}771 &  5.3 \\
\quad & Men                               & 25{,}961 & 1{,}166 & 22{,}617 & 2{,}178 &  4.5 \\
\midrule
\multicolumn{7}{l}{\textit{Disability}} \\
\quad & Physical Disabilities             & 23{,}845 & 3{,}782 & 17{,}068 & 2{,}995 & 15.9 \\
\quad & Visually Impaired                 & 23{,}563 & 2{,}929 & 17{,}788 & 2{,}846 & 12.4 \\
\quad & Hearing Impaired                  & 24{,}183 & 2{,}995 & 18{,}394 & 2{,}794 & 12.4 \\
\quad & Cognitive/Learning Disability     & 24{,}960 & 2{,}618 & 19{,}415 & 2{,}927 & 10.5 \\
\quad & Mental Health Problems            & 25{,}291 & 1{,}732 & 20{,}621 & 2{,}938 &  6.8 \\
\midrule
\multicolumn{7}{l}{\textit{Sexual Orientation}} \\
\quad & Lesbian                           & 22{,}382 & 3{,}191 & 17{,}061 & 2{,}130 & 14.3 \\
\quad & Bisexual                          & 24{,}833 & 3{,}120 & 18{,}681 & 3{,}032 & 12.6 \\
\quad & Gay                               & 22{,}623 & 2{,}552 & 17{,}798 & 2{,}273 & 11.3 \\
\quad & Heterosexual                      & 23{,}830 &    854 & 21{,}432 & 1{,}544 &  3.6 \\
\midrule
\multicolumn{7}{l}{\textit{Origin}} \\
\quad & Migrant Workers                   & 23{,}419 & 5{,}021 & 15{,}519 & 2{,}879 & 21.4 \\
\quad & Immigrants                        & 24{,}088 & 4{,}972 & 15{,}757 & 3{,}359 & 20.6 \\
\quad & Refugees                          & 23{,}971 & 4{,}826 & 16{,}114 & 3{,}031 & 20.1 \\
\quad & Specific Country Origin           & 22{,}170 & 3{,}528 & 15{,}597 & 3{,}045 & 15.9 \\
\quad & Undocumented People               & 21{,}377 & 3{,}020 & 15{,}629 & 2{,}728 & 14.1 \\
\midrule
\multicolumn{7}{l}{\textit{Overall}} \\
\quad & \textbf{All 34 groups}            & \textbf{791{,}283} & \textbf{109{,}788} & \textbf{593{,}893} & \textbf{87{,}602} & \textbf{13.9} \\
\bottomrule
\end{tabular}
\end{table*}

\section{Per-Group Diversity Analysis: \method vs.\ HateBenchSet}
\label{app:diversity_analysis}

Section~\ref{sec:persona_effect} reports that \method achieves higher within-group embedding diversity than HateBenchSet on 32 of 34 identity groups.
This appendix provides the full per-group breakdown and additional analysis.

\subsection{Methodology}

For each of the 34 identity groups shared between \method and HateBenchSet, we sample up to 200 hate-labeled texts per dataset and compute all pairwise cosine distances among their Sentence-BERT embeddings (\texttt{all-MiniLM-L6-v2}). The \emph{within-group diversity} of a dataset for a given group is defined as the mean of these pairwise distances: higher values indicate that texts targeting the same group exhibit more varied phrasing, rhetoric, and style. We also compute three lexical diversity metrics---distinct-1, distinct-2, and distinct-3---defined as the ratio of unique $n$-grams to total $n$-grams across all texts.

\subsection{Per-Group Results}

\begin{figure*}[t!]
\centering
\includegraphics[width=\linewidth]{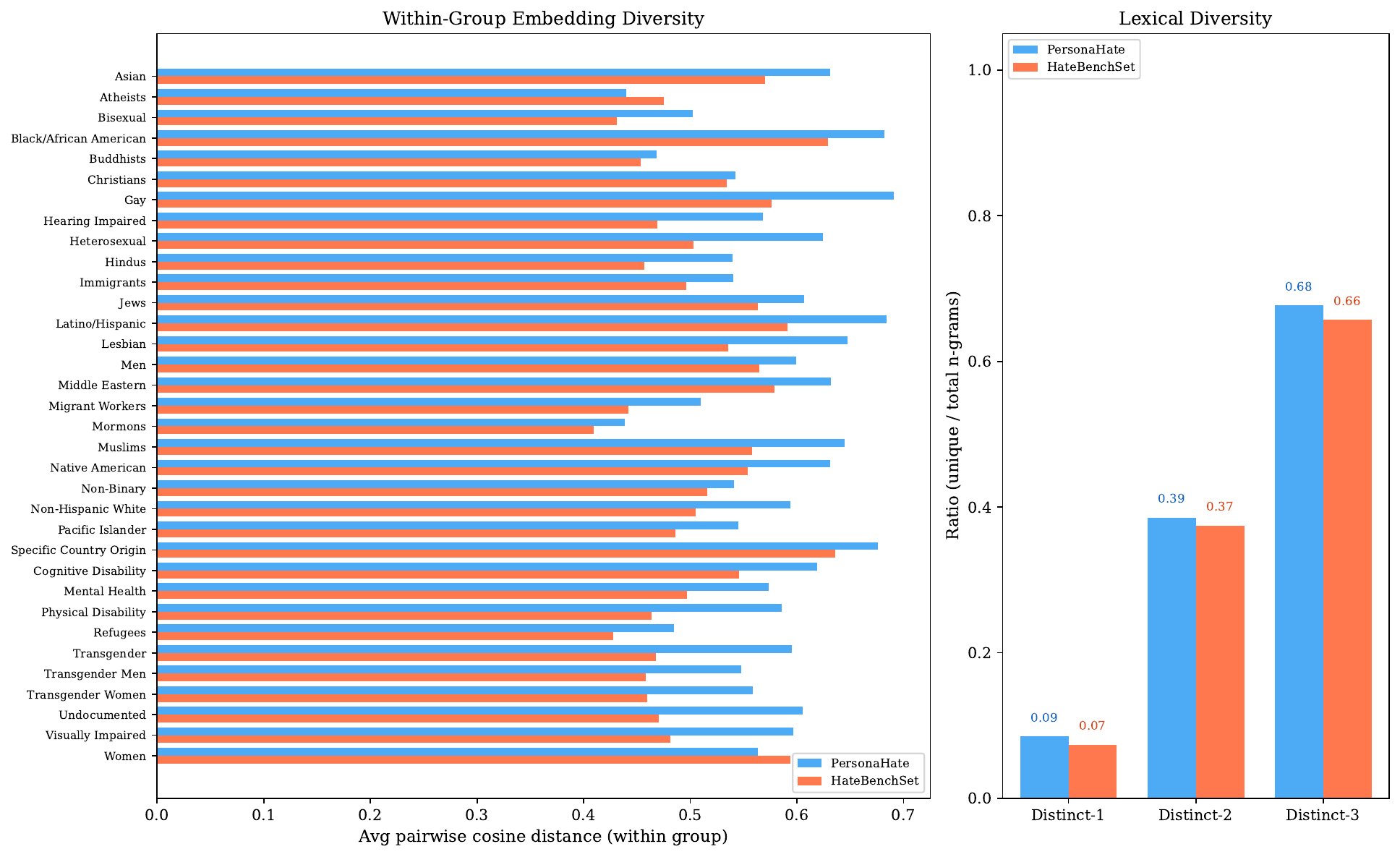}
\caption{\textbf{Left}: within-group embedding diversity (avg pairwise cosine distance among Sentence-BERT embeddings) for all 34 identity groups.
\method shows higher diversity on 32/34 groups. \textbf{Right}: lexical diversity (distinct-$n$) computed over 5{,}000 samples per dataset.
\method achieves higher ratios at all $n$-gram levels.}
\label{fig:diversity_full}
\end{figure*}

Figure~\ref{fig:diversity_full} (left panel) and Table~\ref{tab:within_model_diversity} show the within-group embedding diversity for all 34 groups.
\method (blue) achieves higher diversity than HateBenchSet (orange) on 32 of 34 groups, with the two exceptions being Atheists ($-$0.030) and Women ($-$0.041).
Specifically, \method yields stable within-model diversity across generators (range: 0.449--0.577, std = 0.046), whereas HateBenchSet varies much more widely across models (range: 0.373--0.758, std = 0.131). 
The average within-group cosine distance is 0.580 for \method vs.\ 0.512 for HateBenchSet, a 13.3\% relative improvement.
This suggests that persona conditioning provides a more consistent mechanism for inducing semantic variation, while fixed-template generation is more sensitive to individual model behavior.

Table~\ref{tab:pergroup_diversity} reports the complete numerical results.
The diversity advantage is largest for groups where HateBenchSet outputs are most formulaic: Undocumented (+0.146), Visually Impaired (+0.137), Non-Hispanic White (+0.127), Transgender (+0.126), and Physical Disability (+0.124).
These are groups where direct prompting tends to produce a narrow set of stereotyped outputs, while persona conditioning enables a wider range of rhetorical strategies.

\begin{table*}[t!]
\centering
\caption{Per-group within-group diversity comparison.
Cosine Dist = avg pairwise cosine distance among Sentence-BERT embeddings of hate-labeled texts targeting each group (up to 200 samples).
$\Delta$ = \method $-$ HateBenchSet.
Groups are sorted alphabetically.}
\label{tab:pergroup_diversity}
\small
\setlength{\tabcolsep}{5pt}
\begin{tabular}{lrr|cc|r}
\toprule
\textbf{Identity Group} & \textbf{PH $n$} & \textbf{HB $n$} & \textbf{PH Cosine Dist} & \textbf{HB Cosine Dist} & \textbf{$\Delta$} \\
\midrule
Asian                    & 8{,}753  & 80  & 0.620 & 0.563 & +0.057 \\
Atheists                 & 11{,}021 & 132 & 0.442 & 0.472 & $-$0.030 \\
Bisexual                 & 10{,}786 & 77  & 0.517 & 0.426 & +0.091 \\
Black/African American   & 8{,}335  & 92  & 0.671 & 0.622 & +0.048 \\
Buddhists                & 10{,}824 & 130 & 0.443 & 0.450 & $-$0.007 \\
Christians               & 10{,}988 & 153 & 0.552 & 0.531 & +0.022 \\
Gay                      & 9{,}702  & 84  & 0.686 & 0.569 & +0.117 \\
Hearing Impaired         & 10{,}177 & 87  & 0.581 & 0.463 & +0.117 \\
Heterosexual             & 10{,}834 & 103 & 0.598 & 0.498 & +0.100 \\
Hindus                   & 9{,}215  & 117 & 0.573 & 0.453 & +0.120 \\
Immigrants               & 11{,}075 & 125 & 0.533 & 0.492 & +0.041 \\
Jews                     & 7{,}919  & 101 & 0.571 & 0.558 & +0.013 \\
Latino/Hispanic          & 8{,}962  & 79  & 0.689 & 0.583 & +0.106 \\
Lesbian                  & 9{,}224  & 83  & 0.640 & 0.529 & +0.111 \\
Men                      & 11{,}901 & 147 & 0.594 & 0.561 & +0.033 \\
Middle Eastern           & 8{,}970  & 113 & 0.634 & 0.574 & +0.061 \\
Migrant Workers          & 10{,}456 & 120 & 0.498 & 0.438 & +0.060 \\
Mormons                  & 10{,}243 & 142 & 0.442 & 0.406 & +0.035 \\
Muslims                  & 9{,}002  & 122 & 0.627 & 0.553 & +0.074 \\
Native American          & 9{,}480  & 101 & 0.604 & 0.548 & +0.056 \\
Non-Binary               & 11{,}055 & 79  & 0.547 & 0.509 & +0.037 \\
Non-Hispanic White       & 10{,}349 & 122 & 0.628 & 0.501 & +0.127 \\
Pacific Islander         & 9{,}098  & 94  & 0.580 & 0.481 & +0.100 \\
Specific Country Origin  & 9{,}508  & 100 & 0.669 & 0.629 & +0.039 \\
Cognitive Disability     & 10{,}886 & 99  & 0.643 & 0.540 & +0.103 \\
Mental Health            & 11{,}231 & 125 & 0.561 & 0.493 & +0.068 \\
Physical Disability      & 9{,}841  & 93  & 0.582 & 0.458 & +0.124 \\
Refugees                 & 11{,}041 & 123 & 0.489 & 0.424 & +0.065 \\
Transgender              & 9{,}742  & 95  & 0.589 & 0.463 & +0.126 \\
Transgender Men          & 9{,}462  & 100 & 0.558 & 0.454 & +0.104 \\
Transgender Women        & 9{,}347  & 100 & 0.544 & 0.455 & +0.089 \\
Undocumented             & 9{,}378  & 116 & 0.612 & 0.466 & +0.146 \\
Visually Impaired        & 9{,}812  & 95  & 0.613 & 0.476 & +0.137 \\
Women                    & 11{,}396 & 104 & 0.548 & 0.588 & $-$0.041 \\
\midrule
\textbf{Average}         &     --    &   --  & \textbf{0.580} & \textbf{0.512} & \textbf{+0.068} \\
\bottomrule
\end{tabular}
\end{table*}

\subsection{Lexical Diversity}

Figure~\ref{fig:diversity_full} (right panel) and Table~\ref{tab:diversity_vs_hatebench} compare lexical diversity between the two datasets using 5{,}000 randomly sampled texts each.
\method shows higher distinct-$n$ ratios at all levels: distinct-1 (0.085 vs.\ 0.074, +15\%), distinct-2 (0.385 vs.\ 0.374, +3\%), and distinct-3 (0.677 vs.\ 0.657, +3\%).
The vocabulary size is nearly double: 25{,}129 unique tokens for \method vs.\ 13{,}593 for HateBenchSet at equal sample counts.
This indicates that persona conditioning encourages the use of a broader vocabulary and more varied phrasing, rather than converging on a fixed set of hateful expressions.

\subsection{Analysis of Exceptions}

The two groups where HateBenchSet shows higher diversity---Atheists and Women---are both groups where HateBenchSet has relatively large sample counts (132 and 104 respectively) and where hate speech tends to follow well-established rhetorical patterns across both datasets. For Women, the high diversity in HateBenchSet may reflect the broad range of misogynistic rhetoric captured by its generators (GPT-3, GPT-4, Vicuna, Dolly, OPT), which collectively cover both explicit and implicit forms. For Atheists, the narrow gap ($-$0.030) is within the range of sampling noise given the subsample sizes used.

\section{Persona-Based Prompting Reduces Refusal}

Beyond diversity, persona conditioning also improves the feasibility of generating training data from safety-aligned LLMs. 
Direct requests for hateful content are frequently refused, which can reduce usable yield and skew the dataset toward groups or expressions that pass model safety filters. 
We compare direct prompting and persona-based prompting across three API-gated models: GPT-4o-mini, Gemini-2.5-Flash, and Claude-3-Haiku. 
Each condition uses 100 personas and 34 identity groups, resulting in 3{,}400 prompts per condition per model. 
We detect refusals using a keyword-based classifier covering common refusal phrases such as ``I cannot,'' ``as an AI,'' and ``I apologize.''

As shown in Table~\ref{tab:refusal_cross_model}, direct prompting leads to near-universal refusal across all three models. 
Persona-based prompting substantially reduces refusal rates: GPT-4o-mini drops from 96.2\% to 29.5\%, Gemini from 100.0\% to 79.9\%, and Claude from 100.0\% to 87.7\%. 
This confirms that persona conditioning increases generation yield, although refusal remains substantial for more conservative models. 

\section{Inference Efficiency and Operational Cost}
\label{app:inference_efficiency}

{Table~\ref{tab:inference_efficiency} reports the inference throughput and operational cost of the evaluated models.}
{API models are evaluated with 16 parallel workers, while locally served models are evaluated on a single A100 80GB GPU.}
{Throughput is measured in samples per second (samp/s); the last column reports estimated API cost in USD for API models and GPU-hours for locally served models.}

\begin{table*}[t]
\centering
\caption{{Inference efficiency and operational cost of the evaluated models.
API models use 16 parallel workers, while locally served models are evaluated on a single A100 80GB GPU.
Throughput is measured in samples per second (samp/s).
The last column reports estimated API cost in USD for API models and GPU-hours for locally served models.}}
\label{tab:inference_efficiency}
\small
\setlength{\tabcolsep}{6pt}
\begin{tabular}{llcc}
\toprule
\textbf{Group} & \textbf{Model} & \textbf{Infer (samp/s)} & \textbf{API Cost / GPU-h} \\
\midrule

\multirow{7}{*}{LLM APIs}
& GPT-4o-mini       & 5.2   & \$1.3 \\
& Claude-4-Sonnet   & 5.4   & \$26.8 \\
& Gemini-2.5-Flash  & 13.2  & \$0.5 \\
& Gemini-2.5-Pro    & 1.5   & \$81.7 \\
& Claude Opus 4.6     & 5.4   & \$112 \\
& DeepSeek V4-Flash & 8.3   & \$1.5 \\
& OpenAI Moderation & 19.8  & \$0.1 \\
\midrule

\multirow{4}{*}{Open-source LLMs}
& Llama-3.1-70B     & 16.5  & 1.20 GPU-h \\
& Qwen3-32B         & 47.0  & 0.42 GPU-h \\
& Mistral-Small-24B & 71.7  & 0.28 GPU-h \\
& Llama Guard 3     & 5.2   & 3.80 GPU-h \\
\midrule

\multirow{8}{*}{Encoders}
& BERT              & 3,200 & 0.006 GPU-h \\
& RoBERTa           & 3,500 & 0.006 GPU-h \\
& DeBERTa-v3        & 520   & 0.038 GPU-h \\
& XLM-RoBERTa       & 3,000 & 0.007 GPU-h \\
& Detoxify-BERT     & 3,500 & 0.006 GPU-h \\
& Detoxify-RoBERTa  & 3,200 & 0.006 GPU-h \\
& LFTW-RoBERTa      & 3,000 & 0.007 GPU-h \\
& Cardiff-RoBERTa   & 3,000 & 0.007 GPU-h \\
\midrule

\multirow{4}{*}{Decoder LLMs SFT}
& Qwen2.5-0.5B      & 60    & 0.33 GPU-h \\
& Llama-3.2-1B      & 120   & 0.17 GPU-h \\
& Llama-3.2-3B      & 50    & 0.40 GPU-h \\
& Llama-3.1-8B      & 30    & 0.66 GPU-h \\
\bottomrule
\end{tabular}
\end{table*}

\end{document}